\documentclass[%
 reprint,
 amsmath,amssymb,
 aps,
prstab,
]{revtex4-1}

\usepackage{subfigure}
\usepackage{graphicx}
\usepackage{epstopdf}
\usepackage{dcolumn}
\usepackage{bm}

\begin{document}

\preprint{APS/123-QED}

\title{ Polarization control of an X-ray free electron laser oscillator}

\author{Nanshun Huang$^{1,2}$}

\author{Kai Li$^4$}

\author{Haixiao Deng$^{1,3}$}%
\email{denghaixiao@zjlab.org.cn}

\affiliation{%
$^1$Shanghai Institute of Applied Physics, Chinese Academy of Sciences, Shanghai 201800, China}%
\affiliation{%
$^2$University of Chinese Academy of Sciences, Beijing 100049, China}%
\affiliation{%
$^3$Shanghai Advanced Research Institute, Chinese Academy of Sciences, Shanghai 201210, China}%
\affiliation{%
$^4$Department of physics, University of Chicago, Illinois 60637, USA}%


\date{\today}

\begin{abstract}

High-intensity, fully coherent X-ray radiation with a tunable polarization over a wide spectral range is of great importance to many experiments. In this paper, we propose a tapered crossed-polarized undulator configuration for X-ray free electron laser oscillator (XFELO) to produce arbitrarily polarized X-ray pulses in hard X-ray region. A numerical example utilizing the parameters of the Shanghai High-Repetition-Rate XFEL and Extreme Light Facility (SHINE) is presented to demonstrate the generation of polarization controllable, fully coherent Hard X-ray pulses with 99.9\% polarization degree and 20~KHz polarization switching rate. This scheme also holds the possibility to be used in cavity tunable XFELO.

\begin{description}

\item[PACS numbers] 41.60.Cr

\end{description}
\end{abstract}

\maketitle

\section{\label{sec:introduction} Introduction}

The X-ray free electron laser (XFEL) breaks new ground for X-ray science in biology, chemistry, physics, and the material sciences, for it significantly improves peak photon brilliance as well as the ability to control both the temporal duration and the spectral bandwidth of the produced light pulses compared to synchrotron radiation sources. Linear Coherent Light Source (LCLS) first produced hard X-ray FEL in 2009, after that several XFELs \cite{emma2010first, ishikawa2012compact, Ackermann.2007, Kang2017, milne2017swissfel} have been constructed world widely. In recent years, a large mount of complicated schemes have been developed in pursuit of, e.g., full coherence~\cite{amann2012demonstration, Feng.2018}, multi-color operation~\cite{Marinelli.2015, qiang2011generation}, ultra-fast operation~\cite{Lutman.2016, Zhang.2019}, and large bandwidth operation~\cite{Yan2019, Deng:gb5079}, tunable polarization operation~\cite{Ferrari.2019, lutman2016polarization}.

Tunable polarization is a desirable feature of XFEL user facilities in investigating many essential properties of matter. Traditionally, in soft X-ray region, the circularly polarized laser is able to produce high-resolution Fourier transform holography (FTH) images of magnetic domains~\cite{Eisebitt.2004} and to reveal the ultra-fast dynamics of spin and orbital moments in solids with femtosecond X-ray pulses~\cite{Boeglin.2010}. In hard X-ray region, circularly polarized laser makes it possible to observe inner 3-dimensional magnetic structure with a large dichroic effect for 3$d$ transition metals by employing bulk-sensitive X-ray magnetic circular dichroism measurement~\cite{Toshiya.2017, K_Sato.2010, Fujiwara.2016}. Further, by using the combination of hard X-ray magnetic circular dichroism spectroscopy and Bragg coherent diffractive imaging, one may gain the ability to study magnetic structure and strain mapping in a single X-ray pulse~\cite{Logan.2016, Suzuki.2014}. Purely 45$^\circ$ linearly polarized hard X-ray is also in a considerable demand in quantum-electro-dynamics (QED) experiments on a extreme light station~\cite{Shen.2018}.

However, all free electron lasers based on planar undulators generate only a linearly polarized pulse of a fixed direction (horizontal or vertical). 
In order to obtain the ability to control polarization in XFEL, two kind of schemes have been proposed and applied to most XFEL facilities running in external seeding and self-amplified spontaneous emission (SASE) mode. The first method is to replace planar undulator with elliptically polarized undulator (EPU)~\cite{Temnykh.2008, Sasaki.1994, Spezzani.2011, Allaria.2014}, e.g., APPLE-type and Delta-type undulator, through which electron beam could directly generate circularly polarized radiation. This method is mainly considered to produce circularly polarized soft X-ray, for reasons that its usefulness of lasing in hard X-ray region has not been demonstrated and that the development of such undulators in hard X-ray region will be expensive. In addition, XFELs equipping EPU are unable to fast switch polarization state of the produced radiation, for switching process need slow movement of undulator magnets. Therefor, EPU may not be suitable for a superconducting linac based high-repetition-rate XFELs which naturally enjoy a feature of fast polarization switching. 

\begin{figure*}[!htb]
  \centering
  \subfigure{\includegraphics*[width=480pt]{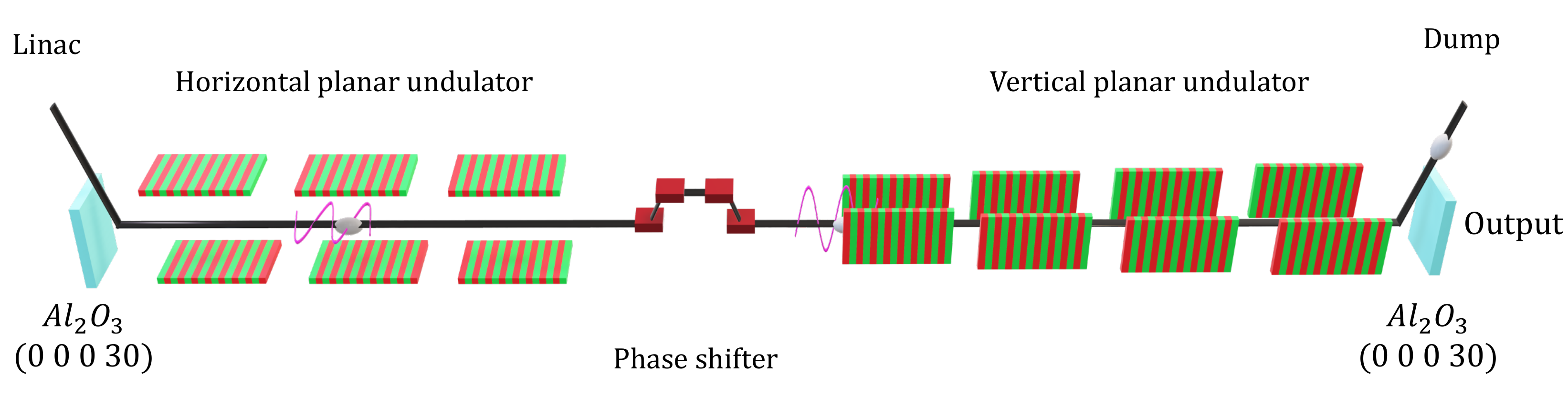}}
  \caption{
   Scheme of tapered crossed undulator scheme in an XFELO using a pair of mirrors. The horizontally polarized radiation field is amplified in first horizontal undulator, and the vertically polarized radiation field is amplified in second tapered vertical undulator by a bunched electron beam. A phase shifter is inserted between the undulators for controlling polarization state. The output radiation is dumped from the downstream mirror.}
  \label{fig:scheme}
\end{figure*}

Another method for generating arbitrarily polarized X-ray is to combine two orthogonally polarized fields produced by crossed-polarized undulators. This scheme contains two planar undulator in a crossed configuration, in which two orthogonally linearly polarized fields could be generated respectively. A phase shifter inserted between the two crossed undulator parts could control the phase difference between the two linear components, resulting in the final polarization state of the combined radiation. The crossed-polarized undulator has been proposed and implemented on synchrotron radiation light sources and FELs~\cite{Kim.1984, Kim.2000}. The advantages of this scheme are the unlimited lasing wavelength and a low cost. However, the polarization switching experiment using crossed-polarized undulator~\cite{Deng.2014} demonstrated that high polarization degree requires longitudinal coherence and stability in intensity. Thus, the significant pulse-to-pulse fluctuations in intensity and spectrum of the SASE XFEL may degrade the polarization performance for generating polarized hard X-ray by crossed-polarized undulator~\cite{Ding.2008}.

XFEL oscillator (XFELO) is one of the candidates for producing stable fully coherent hard X-ray~\cite{kim2008proposal, GainLikai}, with the help of superconducting, MHz-rate linac and high-reflectivity crystal in hard X-ray region. The characteristics of an XFELO pulse, especially enhanced longitudinal coherence and stability compared to SASE XFEL, make crossed-polarized undulator scheme greatly suitable for polarization control in XFELO. The numerical simulations indicate that polarization degree can exceed 99\% with RMS fluctuation of 0.1\%. Furthermore, the crossed-polarized undulator is the best choice that enables XFELO to change the polarization quickly at the present time. With such polarization controllable, fully coherent X-ray, XFELO will significantly improve the capability to generate the single-shot images of spin-resolved electronic structures~\cite{Wang.2012}. 


In this paper, we further the study of a tapered crossed-polarized undulator scheme applied to XFELO for generating fully coherent, arbitrarily polarized X-ray pulses. The paper is organized as follows. In Sec.~\ref{sec:1d solution}, a simple theoretical analysis is performed. In Sec.~\ref{sec:SHINE_example}, we conduct a numerical simulation with the parameters of the Shanghai High-Repetition-Rate XFEL and Extreme Light Facility (SHINE), the first hard X-ray FEL facility in China. Feasibility for using crossed-polarized undulator in a photon energy tunable XFELO is discussed in the following section. Finally, a short summary is given in Sec.~\ref{sec:conclusion}

\section{\label{sec:1d solution} theoretical analysis}

In XFELO, the X-ray pulses circulate in a resonant cavity formed by Bragg reflecting mirrors. In each pass, the pulse is amplified when it overlaps with the next “fresh” electron beam in the undulator, and the spectrum is filtered by the Bragg reflectors. In this process, the relative growth of pulse power per pass can be represented as $G$, and the total reflectivity of mirrors related to the cavity loss and the output can be written as $R_{tot}$. When $(1+G)R_{tot} >1$, a coherent radiation pulse would gradually grows and evolves from the spontaneous radiation. After the exponential growth, the decline of $G$ occurs owing to the electron beam over-modulation by a high-intensity intra-cavity radiation. When $(1+G)R_{tot} = 1 $, the system eventually reaches saturation.

Fig.~\ref{fig:scheme} shows a schematic diagram of the tapered crossed-polarized undulator configuration applied to an XFELO formed by two mirrors. As two undulator sections oriented as right-angle to each other, the two linear polarized components will interact with electron beam in different undulator sections. For horizontally polarized field $E_H$, it is always amplified by a fresh electron beam without the effects came from $E_V$. So that amplification model of $E_H$ can be simplified as an ordinary XFELO using only first horizontal planar undulator section (length $L_H$). Generally, the saturation electron-to-light energy conversion efficiency of an XFELO can be simply evaluated by $ \eta_H = 1/4N_{H}$ where $N_{H}$ is the number of undulator period related to undulator length $L_H$~\cite{Kim.2017, li2017simplified}. This function shows that the saturation energy of $E_H$ is negatively proportional to the undulator length $L_H$. Meanwhile, to guarantee the cavity stability and to growth the radiation field, a larger $L_H$ is preferable. Thus, optimization of the first stage undulator is to balance the cavity stability and output power. 

For vertically polarized field $E_V$, its amplification process would be inevitably influenced by $E_H$ through the bunched electron beam dumped at the end of the first undulator section. At saturation, the dumped electron beam has two main characters: 1) a large bunching factor; 2) a large energy spread. These characters are due to the energy modulation and density modulation of FEL amplification process with high intensity $E_H$ in the first undulator section. To amplify $E_V$ by a bunched electron beam with large energy spread, a post-saturation undulator taper is preferable.  

In Kroll-Morton-Rosenbluth (KMR) model of tapered FEL~\cite{Kroll.1981}, the resonant phase $\psi_{r}$ is given by
\begin{equation} 
  \frac{d\gamma_r(z)}{dz} = - \frac{eK[JJ]E_{V}}{2 \gamma_{r}(z) mc^2} \sin \psi_{r}.
  \label{eq:kmr_model_resonant}
\end{equation}


Then, the energy extraction efficiency of vertical direction can be written as:
\begin{equation}
  \eta_{V} \approx \frac{e}{2m_{e}c^{2} \gamma_{0}^{2}} \int_{0}^{L_{V}} K_V(z)[JJ] f_{\Lambda}(z) E_V(z) \sin \psi_{r}(z) dz
\end{equation}
where $f_{\Lambda}$ is the fraction of trapped electrons, $\gamma_0$ is the electron energy, $\gamma_r$ is the resonant energy, $K_V$ is undulator parameter, $E_V$ is field amplitude of the vertically polarized radiation, $c$ is the light speed, and $m_e$ is the electron mass\cite{Kim.2017}. Here, constant electron energy is assumed, which is a good approximation in low-gain FEL~\cite{Emma.2017}. Taper profile $K_V(z)$ would be optimized ot maximize the energy extraction efficiency. The optimization of the second stage undulator is to decrease undulator length $L_V$ and to make $E_V$ close to $E_H$. 





According to the superposition principle, the total field is given by
\begin{equation}
  E_T = E_{H}+E_{V}e^{i\phi}.
\end{equation}

Here, the relative phase difference $\phi$ determining the final polarization state is separated in two part as $\phi = \phi_{0} + \delta\phi$. The first, $\phi_{0}$, is the generic phase difference between the horizontal and vertical components. The second, $\delta\phi$, is the additional phase modulation which can be varied by tuning the phase shifter between two undulator module. 

According to FEL theory, the radiation phase changes in proportional to the inverse of the radiation amplitude multiplied by $\cos(\psi_{r})$ in tapered undulator~\cite{Prosnitz.1981}. Thus, the generic phase difference $\phi_0$ is always changing until $E_H$ and $E_V$ reaches steady-state in XFELO. The coupled system between $\phi_0$ and $E_V$ raises difficulty in taper optimization, for the variation of $\phi_0$ may cause serious phase mismatch in second undulator section~\cite{Mak.2017, Mak.2015}.









In this case, the polarization properties can be studied using the formalism of Stokes parameters which is defined as~\cite{born_wolf_bhatia_clemmow_gabor_stokes_taylor_wayman_wilcock_1999} :

\begin{equation}
\begin{array}{l}
  {S_{0}=\left|E_{H}\right|^{2}+\left|E_{V}\right|^{2}=I_{H}+I_{V}} ,\\ 
  {S_{1}=\left|E_{H}\right|^{2}-\left|E_{V}\right|^{2}=I_{H}-I_{V}} ,\\ 
{S_{2}=\Re\left(E_{H} E_{V}\right)=I_{45^{\circ}}-I_{135^{\circ}}} ,\\ 
{S_{3}=\Im\left(E_{H} E_{V}\right)=I_{R}-I_{L}}.
\label{eq:stokes_parameters}
\end{array}
\end{equation}

In the functions above, $I_{H}, I_{V}, I_{45^{\circ}}$, and $ I_{135^{\circ}}$ are the intensity for a linearly polarized components over indicated direction, $I_R$ and $I_L$ are the intensities of right-hand and left-hand circularly polarized components respectively.

Using Stokes parameters in Eq.~(\ref{eq:stokes_parameters}), the linearly and circularly polarized fraction of the resulting radiation are
\begin{equation}
  \begin{aligned}
  P_l &=\frac{ S_{2} }{ S_{0} }, \\
  P_c &= \frac{ S_{3} }{ S_{0} }.
\end{aligned}
\end{equation}


Then, polarization states of light can be described, i.e., $P_c=1$ for right-hand circularly polarized, $P_c= -1$ for left-hand circularly polarized, $P_l=1$ for linearly polarized ($+45^\circ$), and $P_l=-1$ for linearly polarized ($135^\circ$). Therefor, the polarization degree can be expressed as $|P_c|$ or $|P_l|$ for circularly and linearly polarized radiation respectively. In addition, we use the on-axis radiation intensity and phase of radiation pulse to calculate the degree of polarization.

\section{\label{sec:SHINE_example} An example design for SHINE}

As a numerical example, we simulated an example XFELO operation with the crossed-polarized undulator configuration for SHINE. As a superconducting accelerator based XFEL, the electron beam will have an energy~8GeV and a 100 pC charge compressed to a peak current of 1000~A with repetition rate of~1MHz. Here, we note that the peak current is lower and bunch length is longer than SASE mode, because a shorter pulse would generate a larger bandwidth resulting in a lower integral reflectivity over spectrum and a larger coupling output in XFELO. So that 1000~A peak current with a longer bunch length is selected. Additionally, SHINE will equip the gap-variable separate undulators each 4~m long with a period length of 16~mm, which can be independently tuned. Phase shifter is to be inserted at the break section between two undulator segments. Additional parameters are listed in Table~\ref{tab:SHINE_para}.

\begin{table}[!htb]
  \centering
\caption{\label{tab:SHINE_para}%
Main parameters of XFELO operation for SHINE.
}
\begin{tabular}{lcr}
\hline
\textrm{Parameter}&
\textrm{Value}\\
\hline
Beam Energy         & 8~GeV    \\ 
Relative Energy Spread     & 0.01\%     \\ 
Normalized Emittance        & 0.4~mm$\cdot$mrad     \\
Peak Current      & 1000~A     \\
Undulator Period Length    & 16~mm     \\
Undulator Segment Length & 4~m \\
Photon Energy     & 14.3~keV     \\
Mirror Material   & Sapphire     \\
Bragg Mirror Reflectivity, $R_{tot}$ & 80\% \\
Darwin Width                           & 13~meV \\
\hline
\end{tabular}
\end{table}

\begin{figure}[!htb]

  \centering
  \subfigure{\includegraphics*[width=240pt]{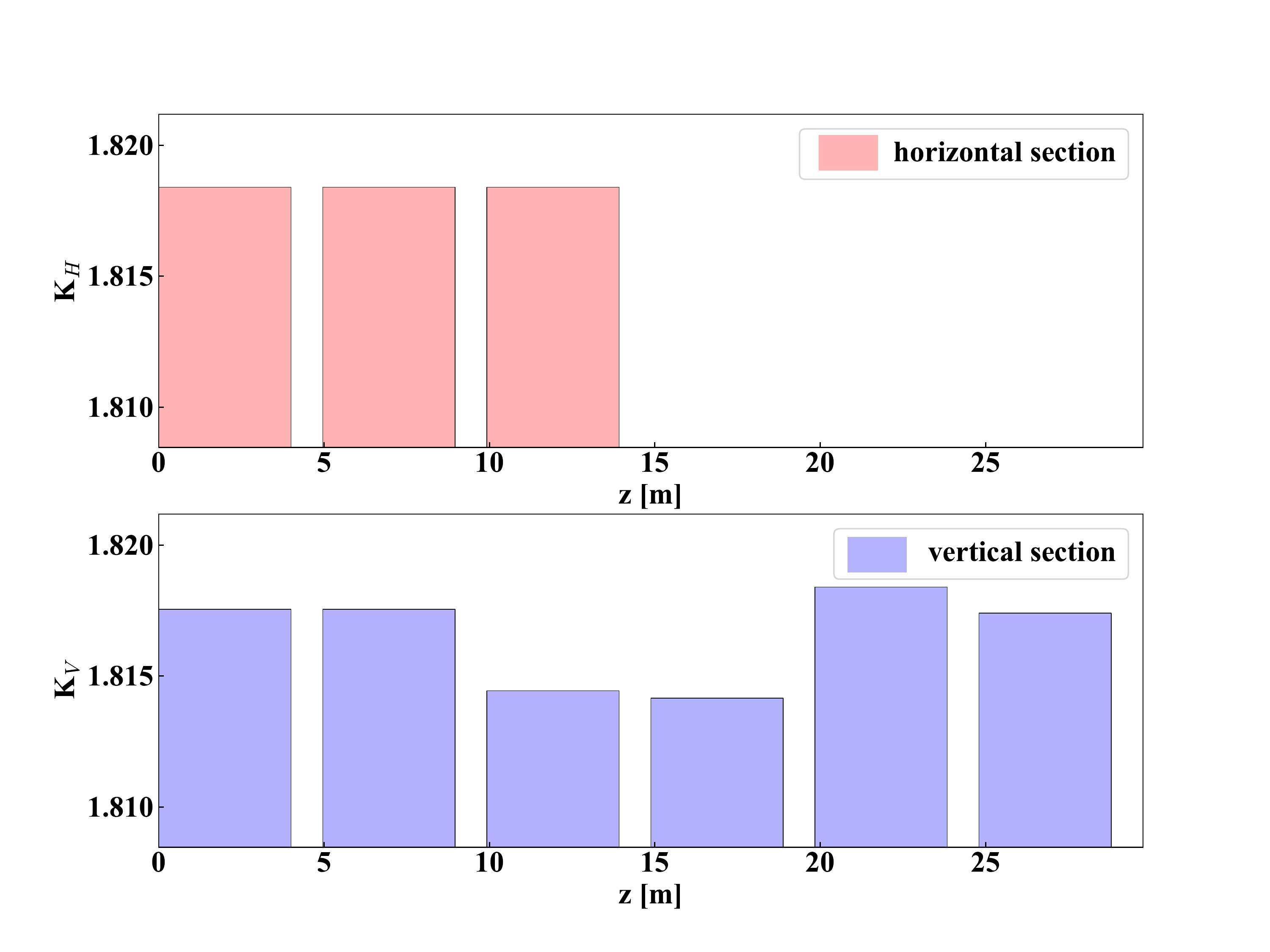}}
  \caption{ The undulator parameter in first undulator section. The optimized taper profile of second undulator section. The undulator parameter of each six undulator is optimized to match the pulse energy produced by first undulator section.}
  \label{fig:und}
\end{figure}

The optical cavity is built from two sapphire (0, 0, 0, 30) crystal mirrors. The length of the cavity reach 150~m so as to match the repetition rate of electron beam. For the whole cavity, the total reflectivity $R_{tot}$ with the Bragg energy of 14.3~keV reaches 80\% (coupling output is 17\%). Meanwhile, the Bragg reflecting mirror is filtering out the frequency components beyond its high-reflectivity spectral bandwidth which is nearly 13~meV. 

In the cavity, nine undulator segments are employed. Three of the undulator segments are horizontal, in order for the horizontally polarized radiation $E_H$ to grow. And the other six segments are vertically placed. Since intensity of $E_V$ is required to be same as it of $E_H$, the undulator parameter $K_V$ of those six segments is carefully optimized by a evolution algorithm. Fig.~\ref{fig:und} present a satisfy taper profile. In this case, five undulator segments would generate a $E_V$ close to $E_H$ but smaller. So six undulator segments are equipped. The last segment can be utilized to precisely control the intensity of $E_V$. In addition, a stair-like taper is better at evaluating the real FEL performance than continuous one. 

\begin{figure*}

  \centering
  \subfigure{\includegraphics*[width=160pt]{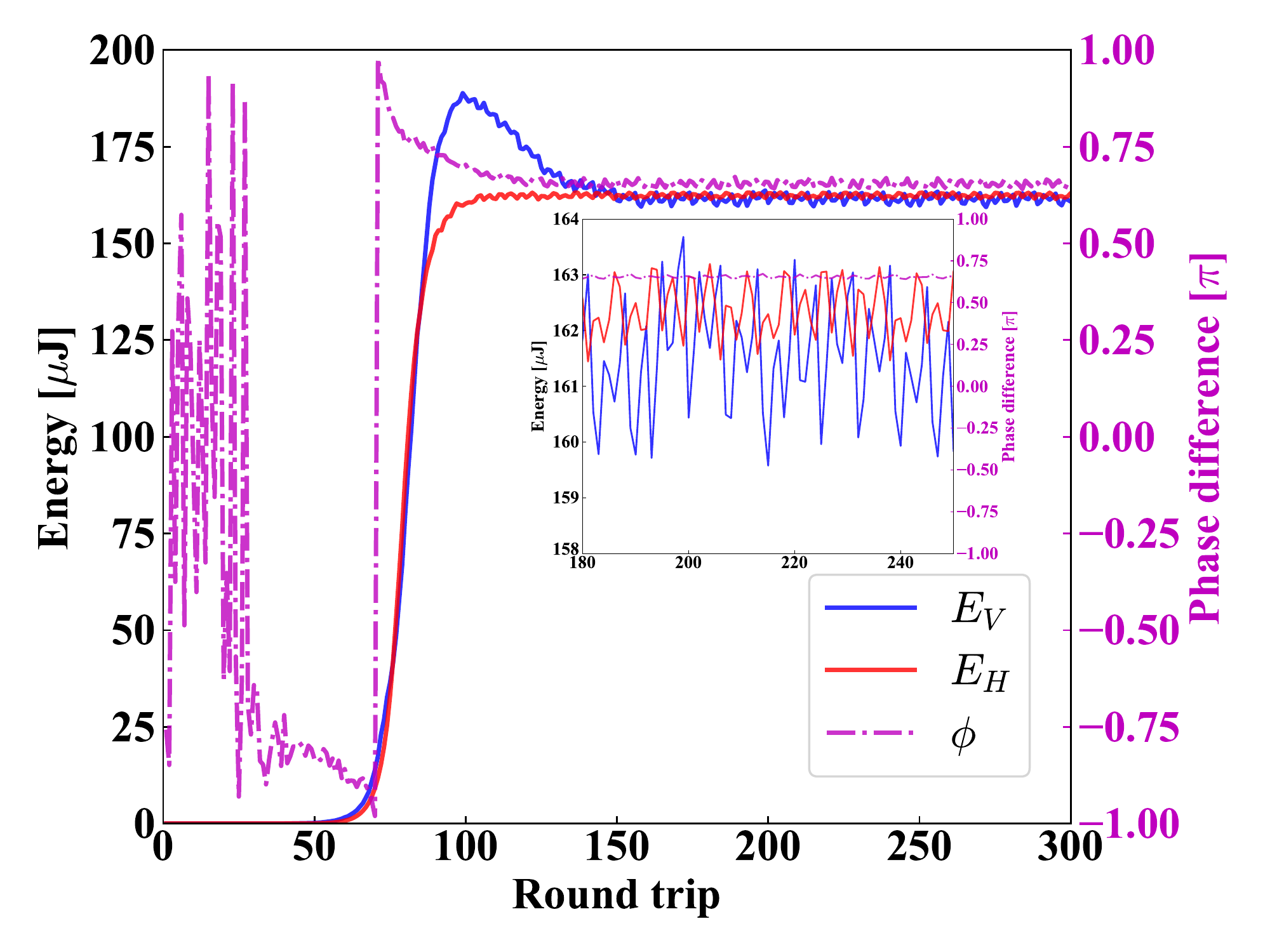}}
  \subfigure{\includegraphics*[width=160pt]{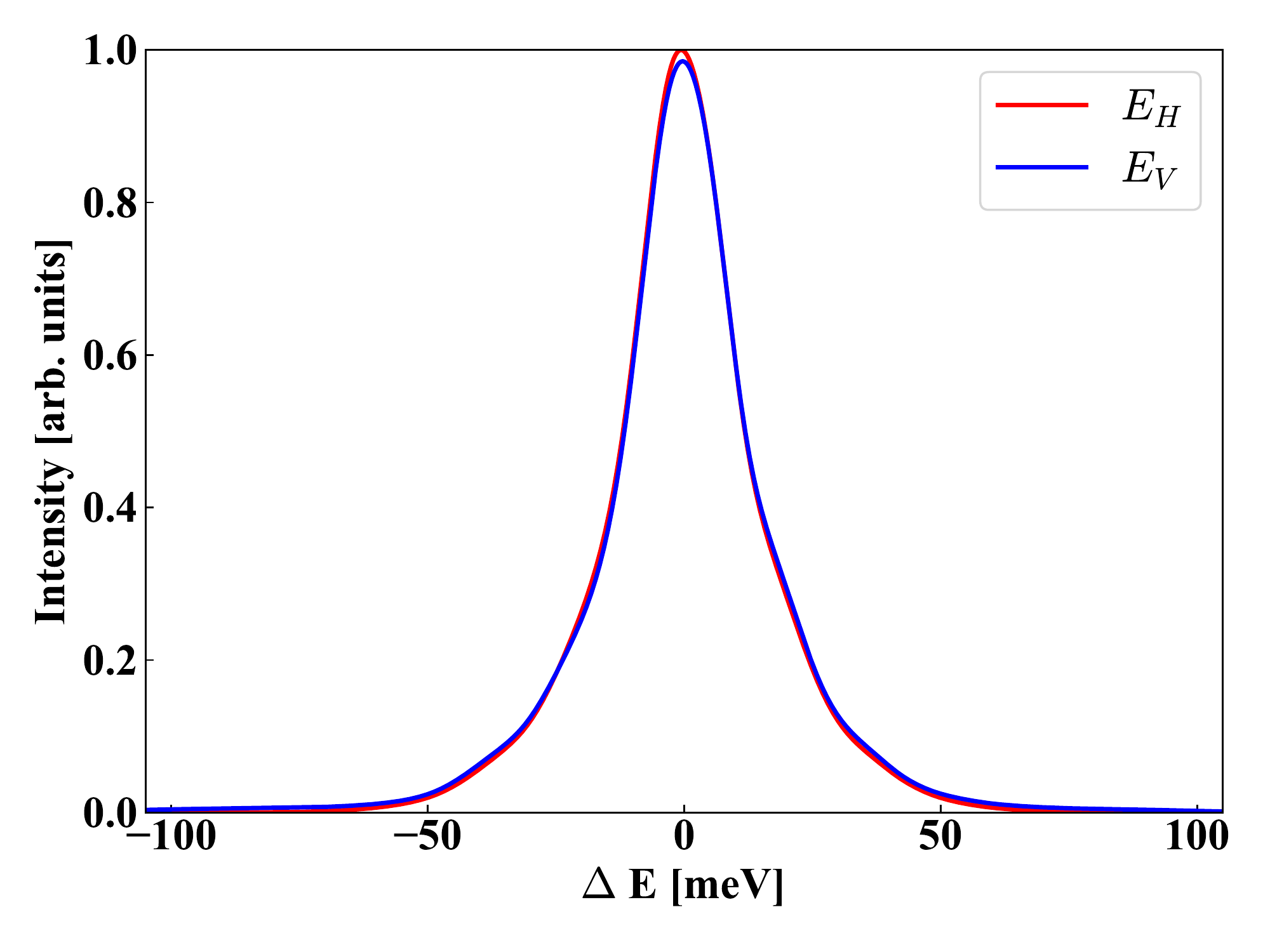}}
  \subfigure{\includegraphics*[width=160pt]{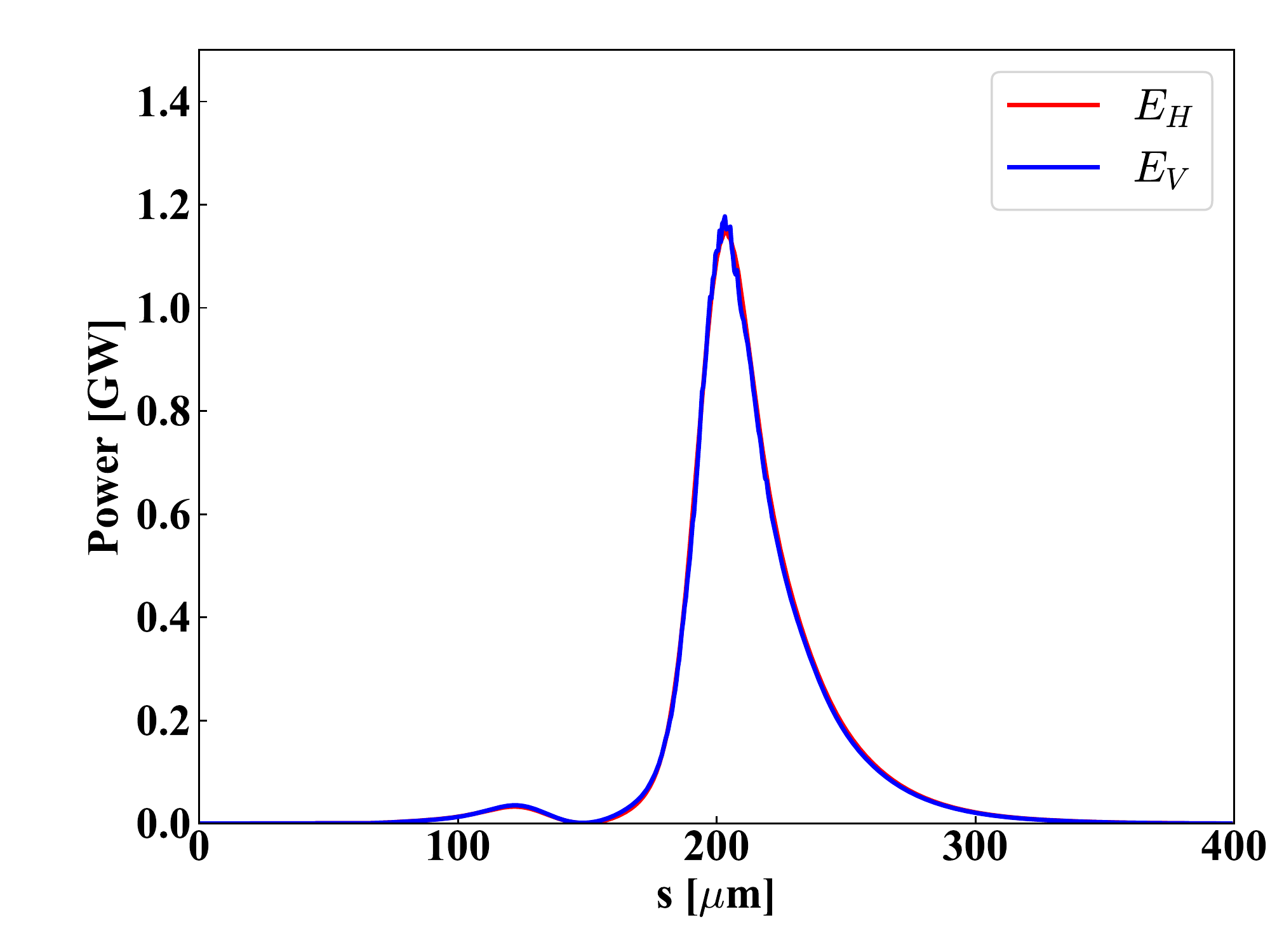}}

  \caption{
   The pulse energy evaluation (left), the spectrum of the radiation (center), and the temporal power profile (right) is shown. The corresponding phase difference between each linearly polarized FEL light (purple line) is presented in the left plot. The pulse has a peak power of 1.2~GW, and the total energy is about 325~$\mu$J. The spectrum has a FWHM width of $25$~meV. }
  \label{fig:evolution}
\end{figure*}

The simulations are conducted by using a combination of time-dependent FEL code GENESIS~\cite{reiche1999genesis}, field propagation simulation code OPC~ \cite{Karssenberg.2006}, and Bragg reflection simulation code BRIGHT~\cite{Huang.2019}. We independently simulate FEL lasing processes in the two sections. The electron beam distribution file is dumped at the end of the first undulator section and is then used as a input in the second undulator.

\begin{figure}

  \centering
  \subfigure{\includegraphics*[width=120pt]{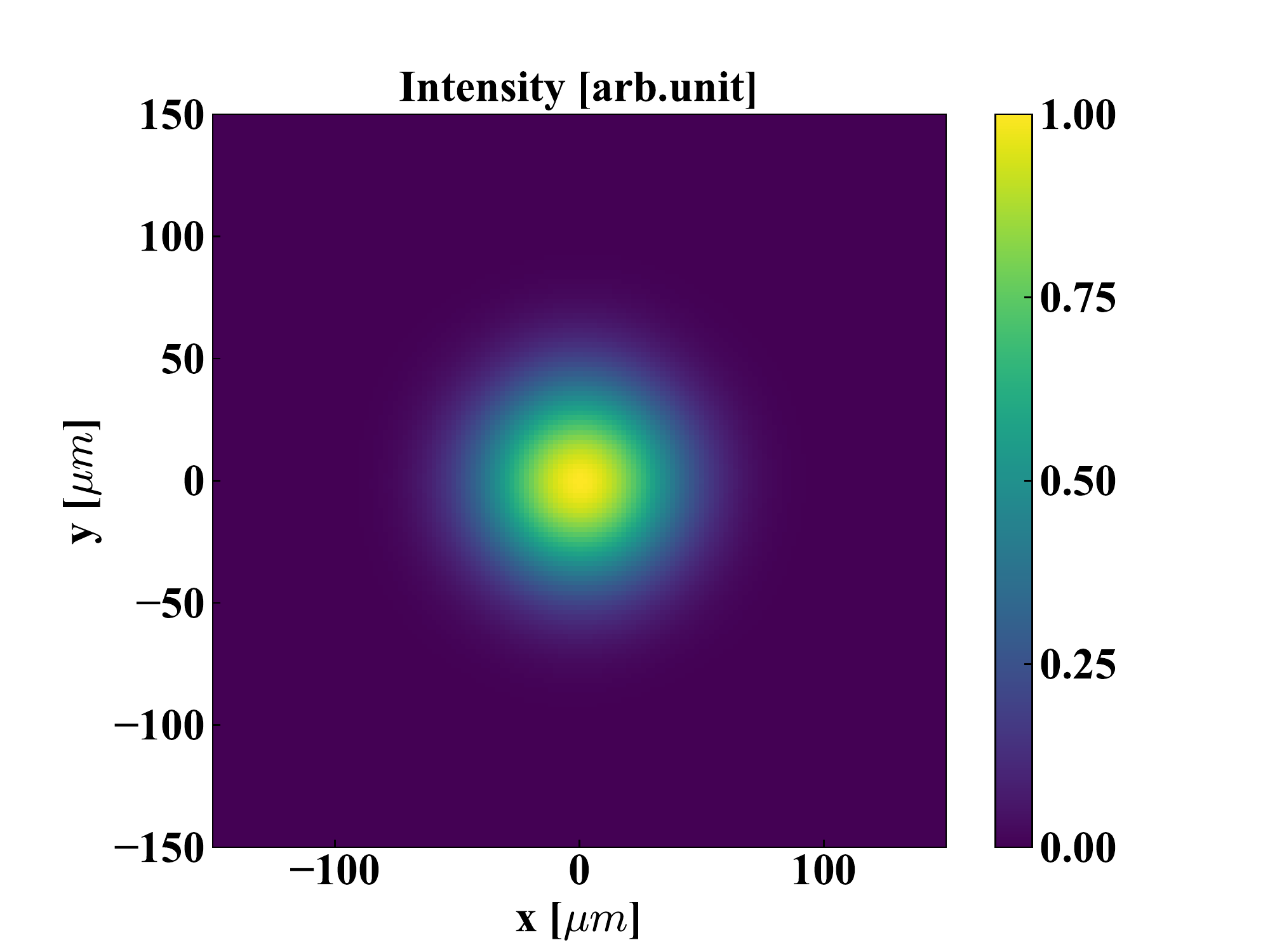}}
  \subfigure{\includegraphics*[width=120pt]{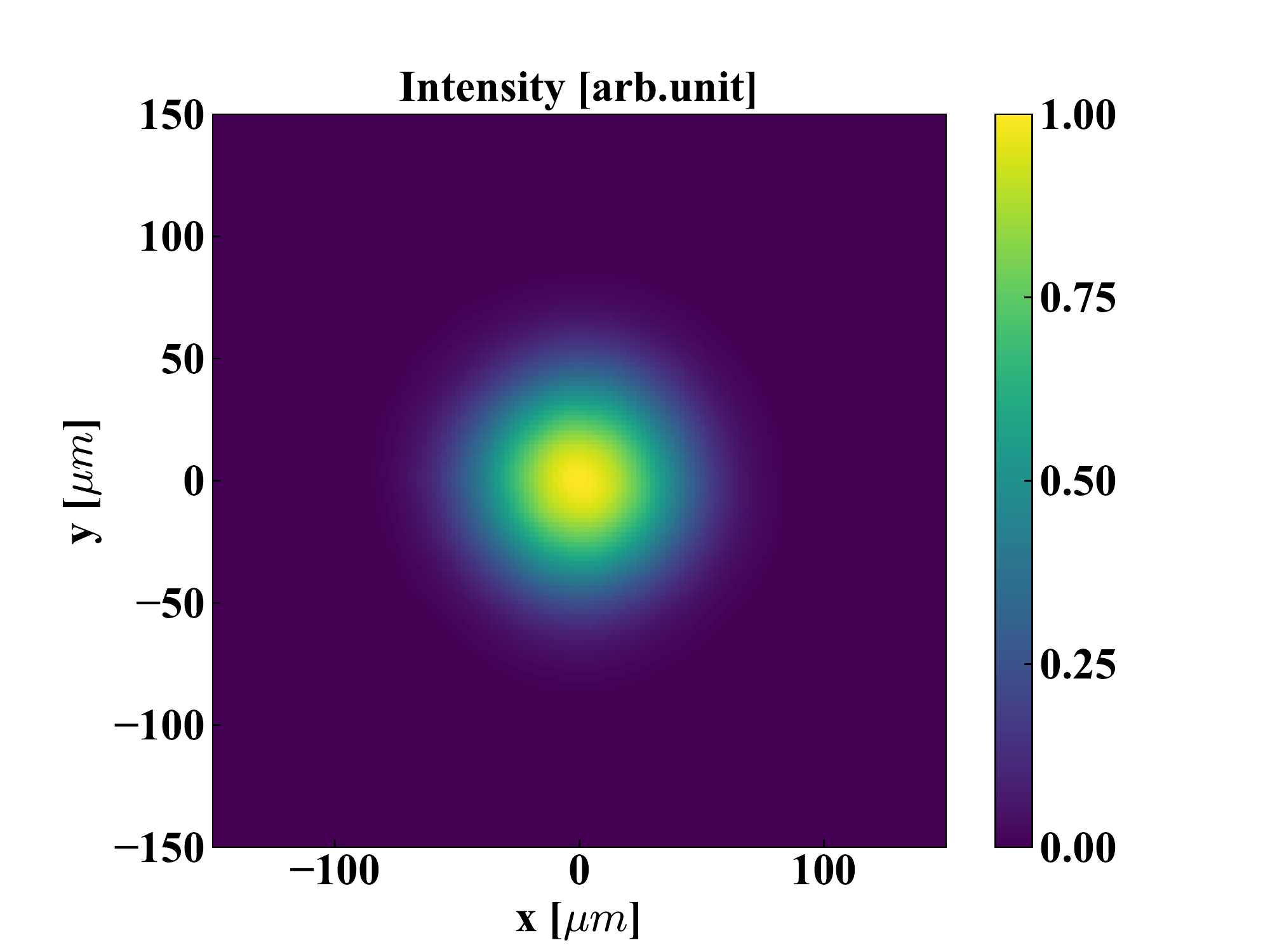}}

  \caption{
    The transverse radiation profile and the transverse phase of $E_H$ (left column) and $E_V$ (right column) are shown. It demonstrates that the outputs have a Gaussian mode.}
  \label{fig:transverse}
\end{figure}

Fig.~\ref{fig:evolution} and Fig.~\ref{fig:transverse} show the simulation results including pulse energy growth, longitudinal characteristics, and transverse mode. The left plot of Fig.~\ref{fig:evolution} shows the pulse energy in both horizontal (red line) and vertical (blue line) direction vs round trip, while the phase difference $\phi$ between them presented in a purple line. After a short period of struggling as the initial shot noise, the laser energy increases as it goes through the undulator again and again. Reached saturation, output pulse energy stays 162~$\mu$j at each polarized direction and total power reach 325~$\mu$j corresponding to $1.4\times 10^{11}$ photons. Thanks to the phase of $E_V$ coupled its intensity and bunching phase of electron beam, which is with related to the phase of $E_H$, $\phi$ varies with pulse energy. The phase difference $\phi$ variation has three stage: irregular variation; regular variation; steady-state with small oscillation. Before exponential growth, the bunching factor is not large enough to influence the phase of $E_V$. So that $E_V$ and $E_H$ grows independently and $\phi$ varies irregularly. As phase of $E_V$ leads by bunching phase, regular variation occurs. After pulse energy saturated, $\phi$ reaches steady-state. The small  $\phi$ oscillation on steady-state results from a instinctive intensity instability of $E_H$ which is a XFELO nature.






The central plot of Fig.~\ref{fig:evolution} presents the spectrum of radiation pulse. The right plot of Fig.~{\ref{fig:evolution}} shows the typical saturation power profile with a peak power of 1.2~GW. The X-ray pulse is a Gaussian-like single peak in both time and spectrum. The the FWHM spectral width is 25~meV, while the temporal width fixed by the current profile to be of about 114~fs. This corresponds to a bandwidth-temporal product of 0.68, approaching the Fourier limit of 0.44 for a Gaussian pulse profile. 

It is the most important considerations in perfect polarization switching of crossed-polarized undulator that the radiation pulses from the two undulator sets should be as close to identical as possible. Fortunately, the negligible slippage effect and the narrow crystal spectral filtering width of XFELO give advantages for this goal. The negligible slippage effect guarantee the generation of two fields with nearly identical profile. The longitudinal coherence resulting from narrow crystal spectral filtering guarantee the coherent superposition of the two components. Thus, the combined radiation is expecting have a excellent polarization performance.







In addition, the transverse radiation profiles are reported in Fig.~\ref{fig:transverse}. The radiation at each direction also holds a nearly identical transverse size. This is because that XFELO transverse mode will be determined by an additional focusing optical elements which is used to control the radiation size and to trap radiation. The nearly identical transverse mode would help to avoid the reduction of polarization when radiation propagate in free space between the source and exponential place.



\subsection{Right-hand and left-hand circular polarization}

\begin{figure}[!htb]

  \centering
  \subfigure{\includegraphics*[width=260pt]{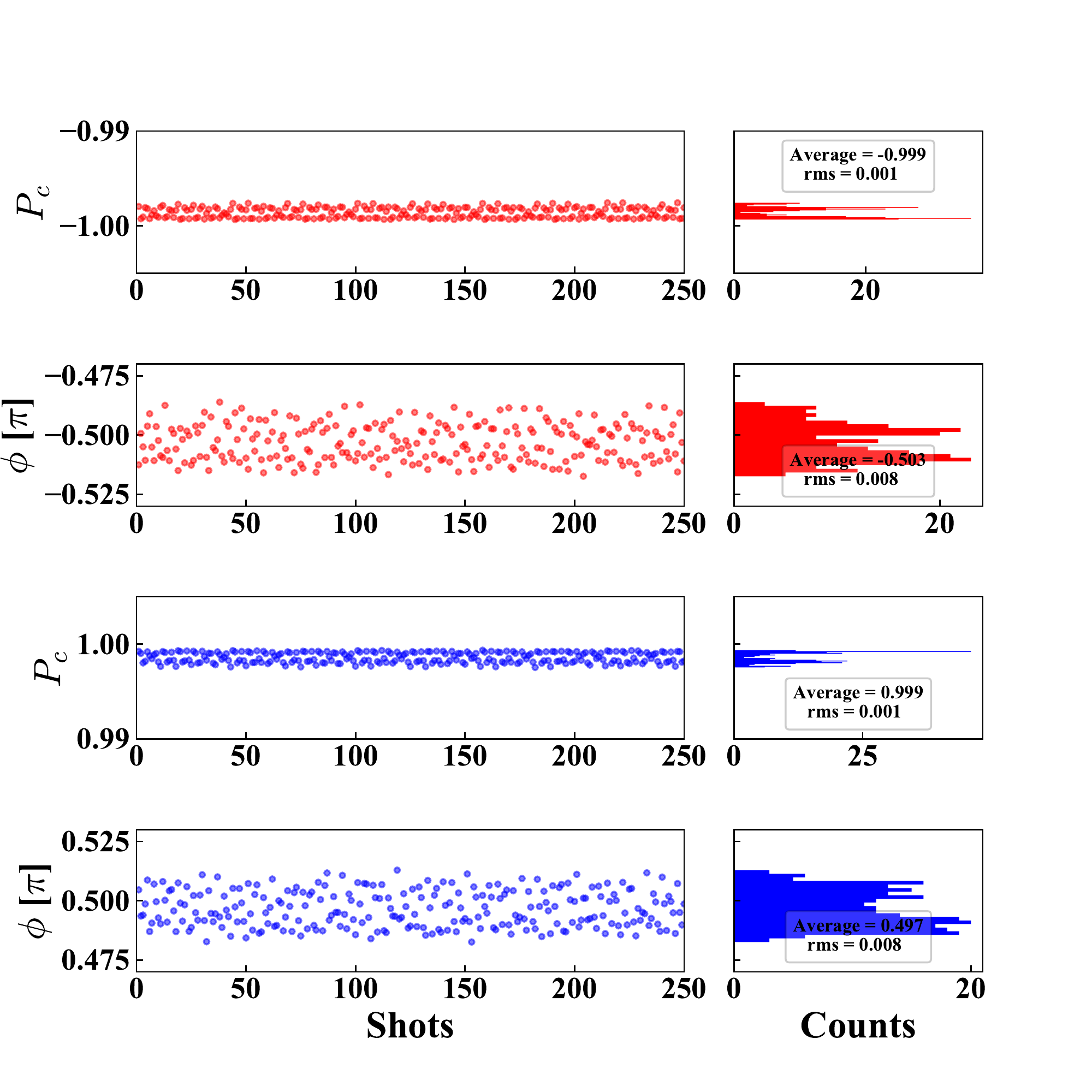}}
  \caption{
    $P_c$ and phase difference $\phi$ when the phase shifter are tuned to left-hand circularly and right-hand circularly. The histograms show the distribution of the data. The average polarization degree $|P_c|$ approach 99.9\% with 0.1\% RMS fluctuation. The histograms also show that the phase difference $\phi$ is obviously stable. }
  \label{fig:circular}
\end{figure}

Benefited from the high stability of the XFELO, the output polarized x-ray beam is expected to be of highly robustness. The statistical result is obtained over 250 shots with different initial random seeds in the GENESIS simulation. The circular polarization performance are presented in Fig.~\ref{fig:circular} in both left-hand (red) and right-hand (blue) polarized state. The average degree of linear polarization exceeds 99.9\% for both polarized state, while the RMS fluctuation levels are 0.1\%. Compared to SASE XFEL, XFELO using crossed undulator could produce a very high degree polarization for left-hand and right-hand. The simulation results confirm that crossed undulator highly relies on a the coherence of two components. The results also suggest that using a comprehensive EPU in XFELO to produce circularly polarized radiation is less attractive. 

Such is a XFELO nature that its output has a small jitter. However, polarization degree jitter is enlarged by strong coupling between the $E_H$ and $E_V$. This is due to that electron beam quality is more sensitive on $E_V$ and that the intensity of $E_H$ would influence the electron beam quality. So that $E_V$ gets a larger intensity jitter than $E_H$ and then the polarization degree jitter of the combined radiation is enlarged. Despite it is enlarged, the polarization jitter is expected to be much smaller than that of SASE XFEL.

\subsection{ 45$^\circ$ linear and 135$^\circ$ linear polarization}



\begin{figure}[!htb]
  \centering
  \subfigure{\includegraphics*[width=260pt]{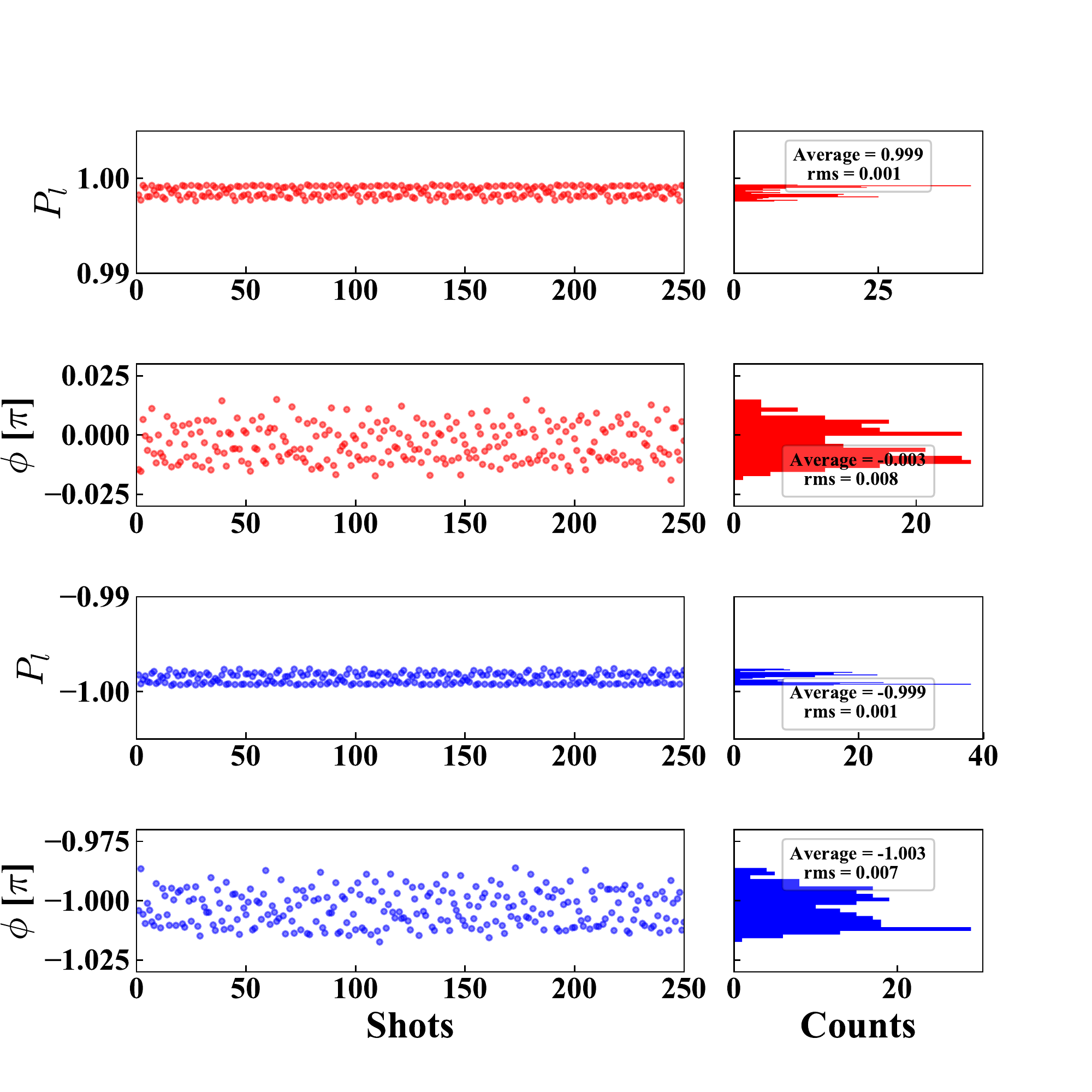}}
  \caption{
    $P_l$ and phase difference $\phi$ when the phase shifter are tuned to $45^{\circ}$ linearly and $135^{\circ}$ linearly. The histograms show the distribution of the data. The average polarization degree $|P_l|$ approach 99.9\% with 0.1\% RMS fluctuation.}
  \label{fig:linear}
\end{figure}

We also calculated the the polarization properties of linearly polarized radiation, again by changing the phase shifter. The results are reported in Fig.~(\ref{fig:linear}) with related to the 45$^\circ$ linearly (red) and 135$^\circ$ linearly (blue) polarized state. The value for each quantity was obtained as an average over 250 shots. The two polarization states reach the same level of polarization degree of 99.9\% and RMS fluctuation 0.1\%, which is similar to the circularly polarized states, see Fig.~\ref{fig:circular}.

In vacuum birefringence experiments, a high intensity linearly polarized X-ray with polarization purity to about $10^{-6}$ are required\cite{Shen.2018}. In addition, the radiation pulse should be initially oriented at an angle of $45^{\circ}$ in the experiment, for the 100~PW laser can not simply change its polarization. Currently, the $45^{\circ}$ linear polarization and its purity is planed to be generated by multiple polarizer, e.g. channel-cut crystal, by the reason that SASE line can not conventionally generate $45^{\circ}$ linearly polarized hard X-ray at SHINE. Even if SASE line generates it, the shot-to-shot intensity fluctuation will be nearly 100\% while monochromator is used to increase polarization purity. In this case, XFELO with crossed-polarized undulator can significantly improve the spectral brilliance with a high stability, for its radiations have fully coherence and high polarization degree.

\subsection{polarization switching}

\begin{figure}[!htb]

  \centering
  \subfigure{\includegraphics*[width=240pt]{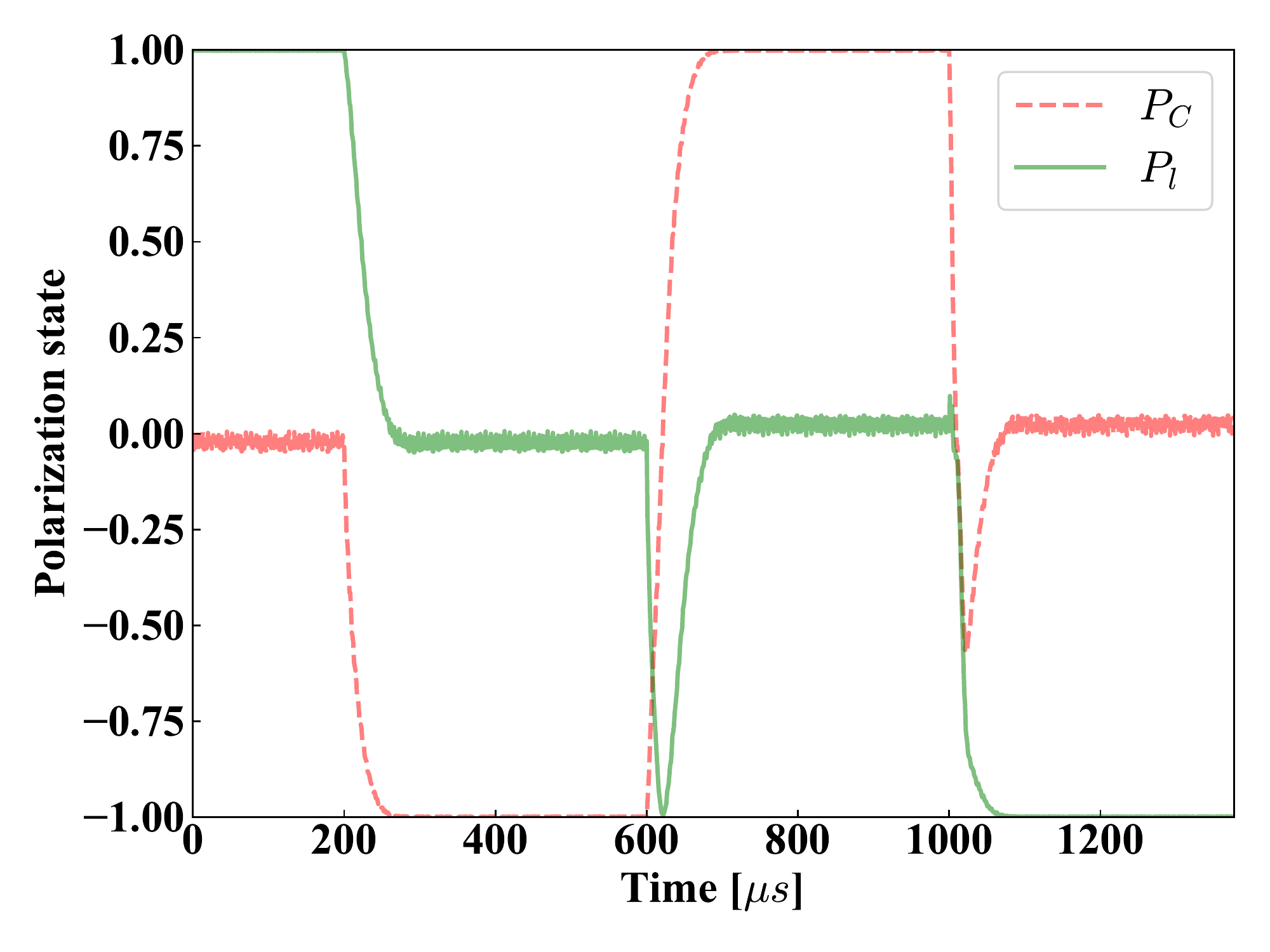}}
  \caption{
    The evolution of the polarization state. $P_c$ displays by red dash line and $P_l$ displays by green line. After changing the phase shifter, phase difference becomes stable within about 50 passes. }
  \label{fig:polarization_degree}
\end{figure}

To demonstrate the performance of polarization switching, the phase shifter is adjusted, which changes the polarization from $45^{\circ}$ linear to left-handed circular, then to right-handed circular, finally to $135^{\circ}$ linear. The calculated $P_c$ and $P_l$ are given in the Fig.~\ref{fig:polarization_degree}. The polarization switching is accomplished in about 50 passes, 50 microseconds for 1~MHz repetition rate, which is four orders of magnitude faster than those requiring mechanical adjustments of undulator magnets. 

It is not like a single pass high-gain XFEL, in which the phase of $E_V$ will be changed immediately in next shot, that vertical field $E_V$ in XFELO will regenerate with the new phase in the next tens round trip, once the phase shifter changed phase. This is an obvious disadvantage of XFELO in comparison with SASE XFEL. A possible way to accelerate the polarization switching is to increase the single pass gain. As long as the single pass gain is large enough, the regeneration of $E_V$ could be accomplished in few round-trip. 

The phase shifter change the longitudinal phase of electron beam typically by induce an additional travelled distance of electron. To achieve this, a small magnetic chicane usually is used as phase shifter. The extra distance delay induced by a magnetic chicane is 

\begin{equation}
  \Delta z = \left| \frac{R_{56}}{2} \right|.
\end{equation}

The delayed distance just few radiation wavelength. Thanks to that the required $R_{56}$ is so small and that the bunching factor could reach 0.2, the effect on bunching factor could be negligible.

\section[]{Discussion For photon energy tunable XFELO}

\begin{figure}[!htb]
  \centering
  \subfigure{\includegraphics*[width=240pt]{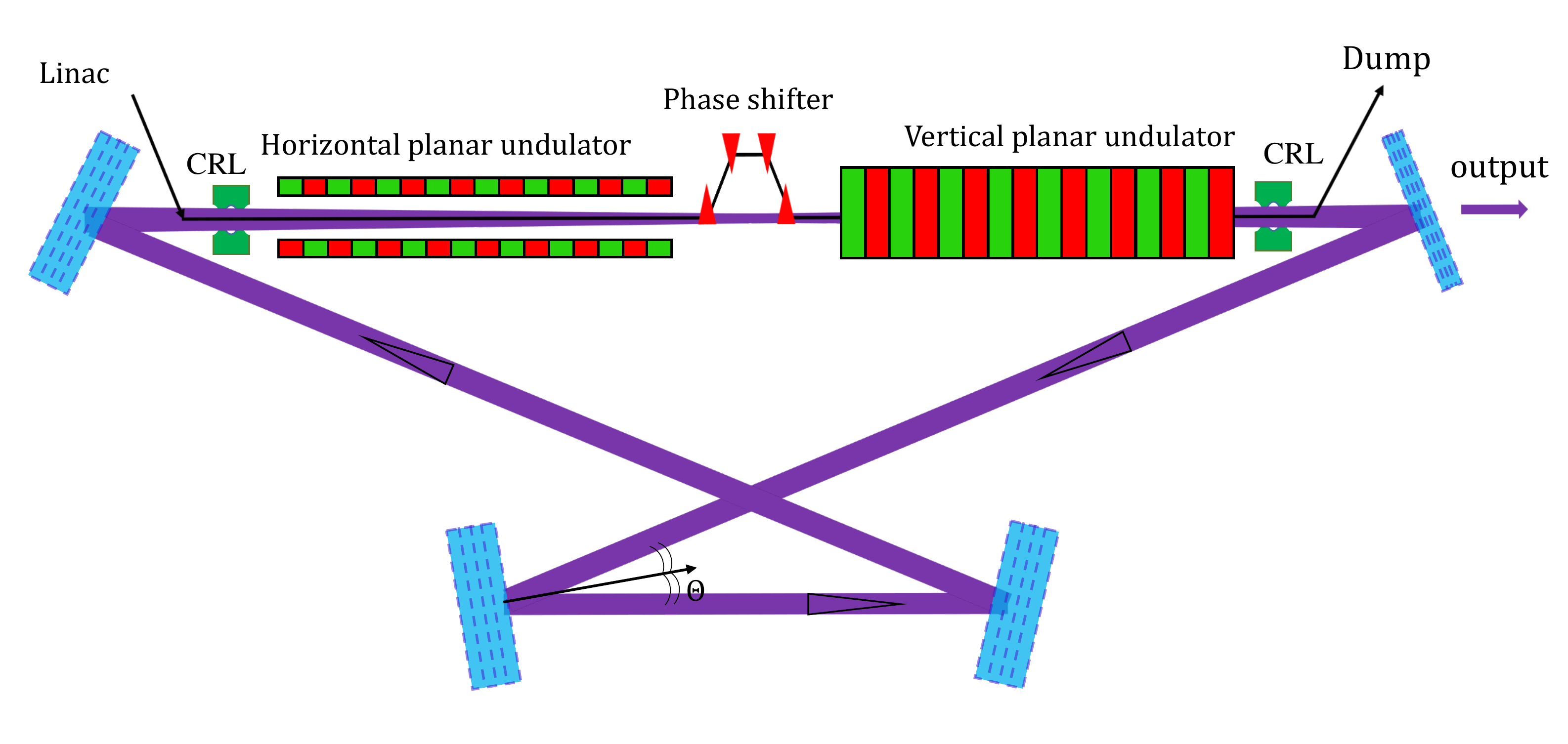}}
  \caption{
   A schematic configuration for the tunable cavity. Four Bragg mirrors are used. When changing the photon energy, the incidence angle of all four mirror must change simultaneously. }
  \label{fig:tunable_cavity}
\end{figure}

The resonant cavity for the energy tunable XFELO consists of four crystal mirrors~\cite{Kim.2017, Lindberg.2011}. The photon energy is tuned by adjusting the X-ray angle of incidence on the four mirrors, see in Fig.~\ref{fig:tunable_cavity}. The range of tunable energy is 

\begin{equation}
  E_{H} \leq E \leq E_{H} / \cos \Theta_{\max }
\end{equation}
where $\Theta = \pi/2 - \theta_B $, $E_{H}$ is Bragg energy and $\theta_B$ is Bragg angle, the angle between the incident optical axis and the crystal plane, shown in Fig.~\ref{fig:tunable_cavity}, $\Theta_{\max }$ is maximum tunable angle. Generally, $\Theta_{\max}$ is less than $20^{\circ}$, so that the tuning energy range for 14.3~keV Bragg energy is about 917~eV. For the tunable cavity using crossed-polarized undulator to generate arbitrary polarization, one should account for the fact that the radiation at different polarized direction has non-equal phase shift and non-equal reflectivity at non-normal incidence. 

\begin{figure}[!htb]

  \centering
  \subfigure{\includegraphics*[width=240pt]{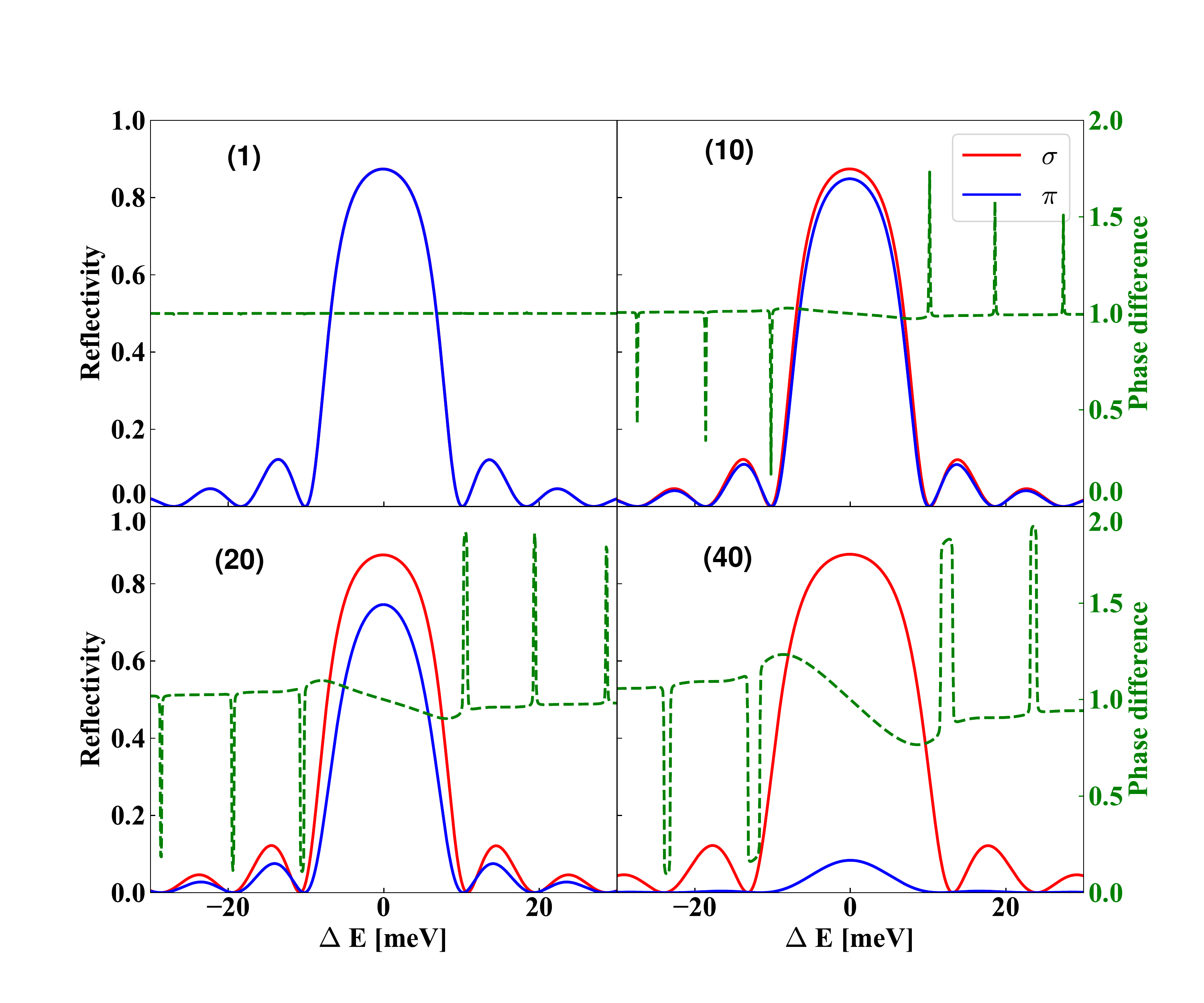}}
  \caption{
    The rocking curves with the additional phase difference between the horizontally polarized and vertically polarized components as a function of relative photon energy $\Delta E$. The calculated results based on the dynamical theory of X-ray diffraction.  $\Theta$ is $1^{\circ}$, $10^{\circ}$, $20^{\circ}$, and $40^{\circ}$}
  \label{fig:phase_angle}
\end{figure}

As mentioned above, Bragg mirror could introduce a extra phase difference between the horizontally polarized and vertically polarized components. Since the angular divergence of the XFELO beam is sufficiently smaller than the Darwin width, we only considered the effect upon relative energy $\Delta E$. Fig.~\ref{fig:phase_angle} shows the the reflectivity on different polarization direction and resulting phase difference (green line), while $\Theta$ is $1^{\circ}$, $10^{\circ}$, $20^{\circ}$, and $40^{\circ}$. The reflectivity is calculated with Diamond (3 3 7) reflection; $E_H$ is 14.225~keV. As the XFELO pulse is not purely monochromatic, phase shift may be significantly different for large $\Delta E$ at the edge of Darwin width. However, there is a limited effect because most of the spectral intensity is pretty contained to the center of Darwin width and the intensity in the central of Darwin width would relocate in temporal distribution with small. Thus, the high weighted average polarization degree could be maintained after Bragg reflection. Further, the crystal mirrors in a dispersive arrangement, i.e., fully symmetric $(+ - - +)$ setting, may cancel the phase variation and transfer the circularly polarized radiation in cavity~\cite{Ishikawa.1989, Malgrange.1991}. 

\begin{figure}[!htb]
  \centering
  \subfigure{\includegraphics*[width=240pt]{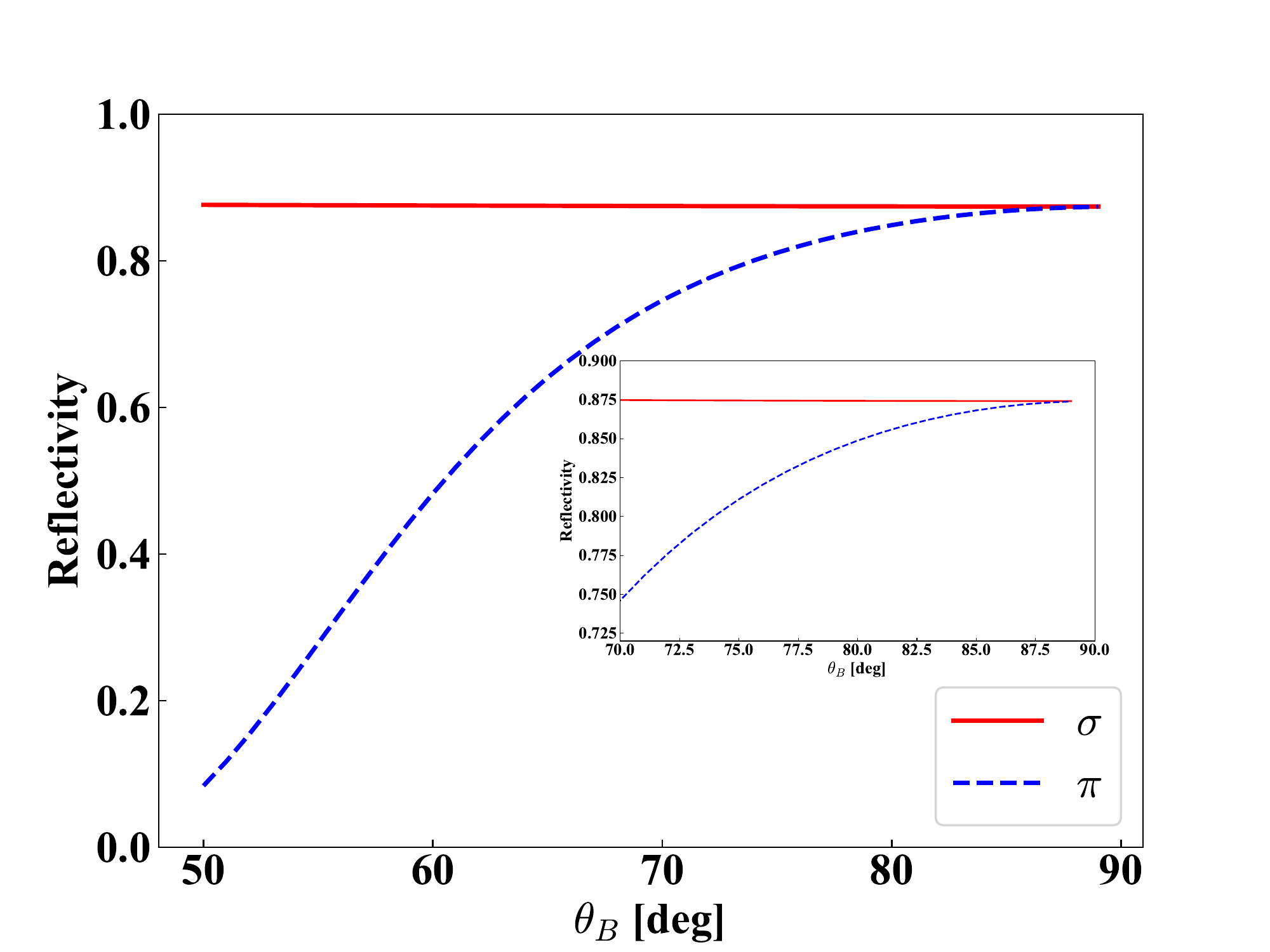}}
  \subfigure{\includegraphics*[width=240pt]{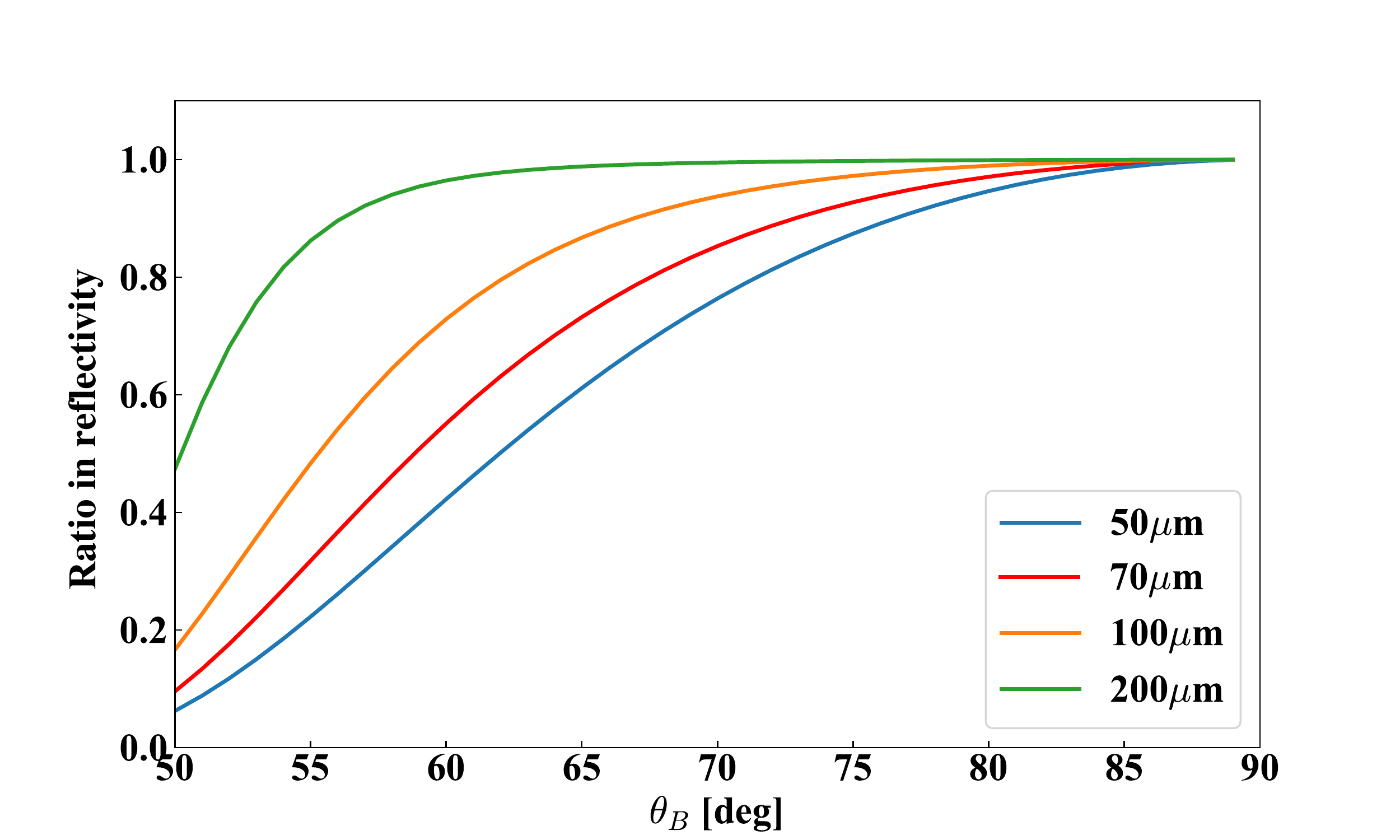}}
  \caption{
   The top plot shows the reflectivity in different polarized direction at different $\theta_B$. The bottom plot shows the ratio of reflectivity in two polarizations with different crystal thickness. The ratio defines as the reflectivity in $\pi$ polarization divides by reflectivity in $\sigma$ polarization.  }
  \label{fig:Ref_d}
\end{figure}

In the top plot of Fig.~\ref{fig:Ref_d} shows the reflectivity on different polarization at $\Theta$ in range from $1^{\circ}$ to $40^{\circ}$. The reflectivity difference between horizontally polarized and vertically polarized components is small, when $\Theta$ is small. While $\Theta$ is smaller than $20^{\circ}$, the ratio in the Bragg reflectivity of the two polarizations could exceed 0.85, which is acceptable and controllable. The bottom plot of Fig.~\ref{fig:Ref_d} displays the reflectivity ratio of two components with different crystal thickness. A 200 $\mu$m thick crystal will generate nearly identical reflectivity while $\theta_B$ larger than $60^\circ$. However, using a thick crystal, output coupling would be a problem. New method to dump radiation out the cavity need to be developed, e.g., a additional thin crystal to extract a part of radiation~\cite{Shvydko2019}. At this situation, the intensity error due to the non-equal reflectivity can be compensated to a negligible degree by tuning the gain of each direction, e.g. by tuning the taper. Thus, non-equal reflectivity effect may not preventing the scheme from working, but limit the tunable Bragg angle and the saturation power.

\begin{figure}[!htb]

  \centering
  \subfigure{\includegraphics*[width=240pt]{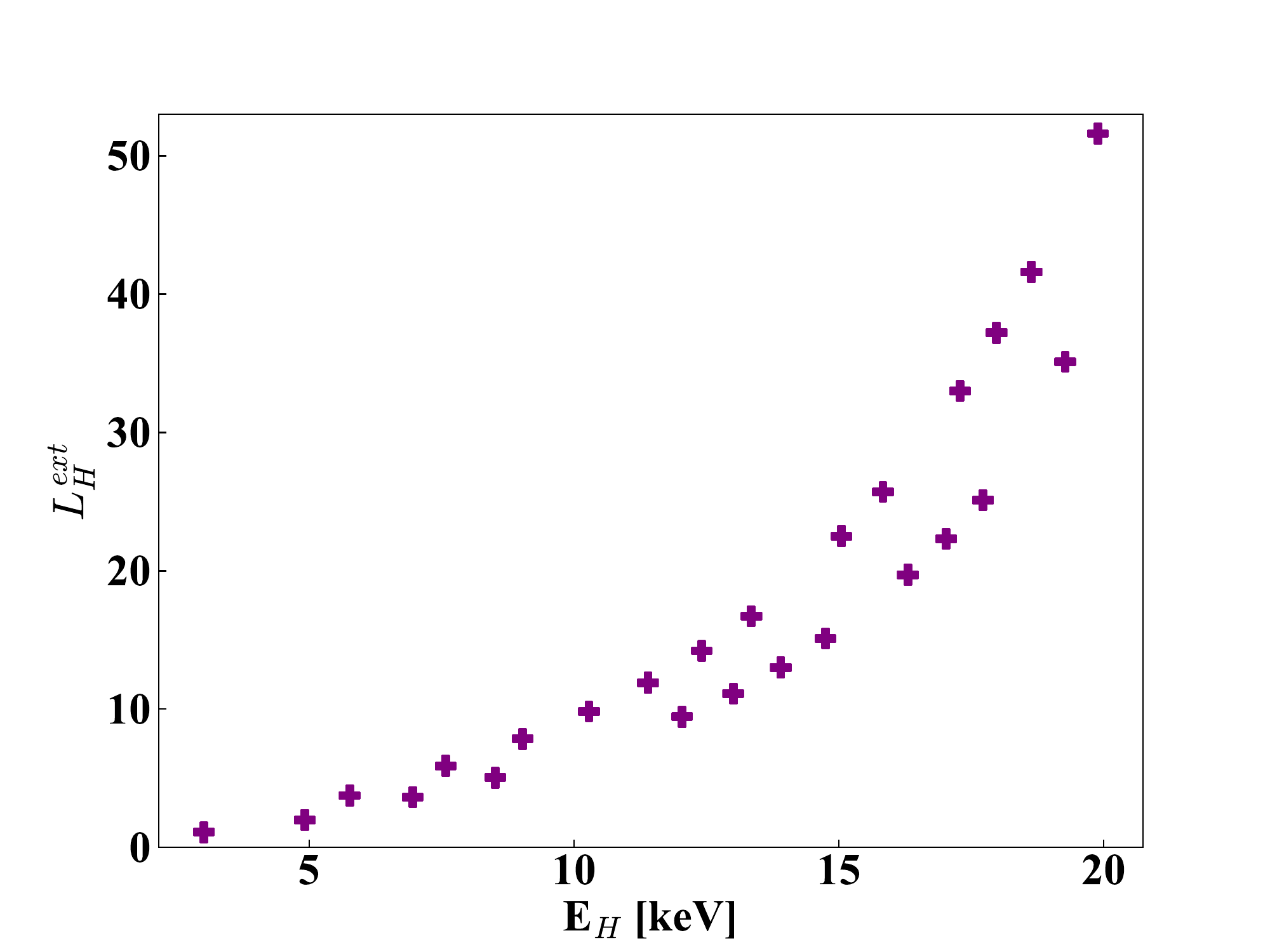}}
  \caption{ The extinction length in various Bragg energy of different reflecting plane.}
  \label{fig:ref_Vs_EH}
\end{figure}

Additionally, Bragg crystal mirror would produce an extra delay attributed to the time required for the radiation to penetrate crystal~\cite{Lindberg.2011}. The penetration distance is related to the crystal extinction length which can be expressed as 

\begin{equation}
  L_{H}^{\mathrm{ext}} =\frac{\lambda_{H}}{ 2 \pi P \left|\chi_{H}\right|}
  \label{eq:extinction_lenght}
\end{equation}
where $\lambda_{H}$ is the wavelength of Bragg reflection, $\chi_{H}$ is Fourier coefficients of the crystal electric susceptibility. $P$ is the polarization factor, for the $\sigma$-polarization component is $P=1$, and $P = \cos 2\theta_B$ for the $\pi$-polarization component~\cite{ShvydKo2012}. Thus, the ratio in the extinction length of the two polarizations is $|\cos 2\theta_B|$. When $\theta_B$ close to $45^{\circ}$, the ratio would be extremely small, which means horizontally polarized and vertically polarized components will have a highly different delay caused by Bragg diffraction. For short pulse it may become a fatal disease. The extinction length in the symmetric scattering geometry varies as Bragg energy $E_H$ of Diamond on different reflection plane is shown in Fig.~\ref{fig:ref_Vs_EH}.

Generally, $L_{H}^{\mathrm{ext}} \approx$ 1-50 $\mu \mathrm{m}$ and it increases with Bragg Energy $E_H$. While $\theta_B$ is $70^{\circ}$, $|\cos 2\theta_B| = |\cos 70^{\circ}|=0.77$ and the difference of delayed distance is 0.92-46 $\mu \mathrm{m}$. Therefor, for a large photon energy and a small $\theta_{B}$, one should increase the bunch length to reduce the influence of different time delay.

\section{\label{sec:conclusion} conclusion}

In this paper, we exploit the taper crossed-polarized undulator scheme for XFELO device in order to achieve the ultimate performance in polarization controlling, yielding the high degree of polarization with the arbitrary state, and fast switching. The undulator line is divided into two sections, and the taper profile of the second undulator is optimized to match the radiation power generated in the first undulator section. Theoretical analysis of energy extraction efficiency of two undulator sections is conducted.

We presented an illustration of the scheme for the SHINE. The proposed scheme can generate fully coherent X-ray pulses with 99.9\% polarization degree and 0.1\% RMS fluctuation . The polarization state can be varied arbitrarily by using a phase shifter inserted between the two undulator sections according to the needs of the user. The main advantage of this scheme appears to be high polarization degree in hard X-ray region with fast polarization switching, up to dozens-KHz level, which is four orders of magnitude faster than those requiring mechanical adjustments of EPU. The possibility for using this scheme in cavity tunable XFELO is discussed. The discussions also show that this scheme could be used in cavity tunable XFELO with limited the energy tune range, limited Bragg angle, and limited electron beam bunch length. The proposed method will provide a powerful tool for polarization experiments at X-ray FEL oscillator facilities. 




\begin{acknowledgments}
The authors are grateful to J.~Yan for helpful discussions and useful comments. This work was partially supported by the National Key Research and Development Program of China (2018YFE0103100, 2016YFA0401900) and the National Natural Science Foundation of China (11775293).
\end{acknowledgments}

\nocite{*}



\begin{thebibliography}{50}%
  \makeatletter
  \providecommand \@ifxundefined [1]{%
   \@ifx{#1\undefined}
  }%
  \providecommand \@ifnum [1]{%
   \ifnum #1\expandafter \@firstoftwo
   \else \expandafter \@secondoftwo
   \fi
  }%
  \providecommand \@ifx [1]{%
   \ifx #1\expandafter \@firstoftwo
   \else \expandafter \@secondoftwo
   \fi
  }%
  \providecommand \natexlab [1]{#1}%
  \providecommand \enquote  [1]{``#1''}%
  \providecommand \bibnamefont  [1]{#1}%
  \providecommand \bibfnamefont [1]{#1}%
  \providecommand \citenamefont [1]{#1}%
  \providecommand \href@noop [0]{\@secondoftwo}%
  \providecommand \href [0]{\begingroup \@sanitize@url \@href}%
  \providecommand \@href[1]{\@@startlink{#1}\@@href}%
  \providecommand \@@href[1]{\endgroup#1\@@endlink}%
  \providecommand \@sanitize@url [0]{\catcode `\\12\catcode `\$12\catcode
    `\&12\catcode `\#12\catcode `\^12\catcode `\_12\catcode `\%12\relax}%
  \providecommand \@@startlink[1]{}%
  \providecommand \@@endlink[0]{}%
  \providecommand \url  [0]{\begingroup\@sanitize@url \@url }%
  \providecommand \@url [1]{\endgroup\@href {#1}{\urlprefix }}%
  \providecommand \urlprefix  [0]{URL }%
  \providecommand \Eprint [0]{\href }%
  \providecommand \doibase [0]{http://dx.doi.org/}%
  \providecommand \selectlanguage [0]{\@gobble}%
  \providecommand \bibinfo  [0]{\@secondoftwo}%
  \providecommand \bibfield  [0]{\@secondoftwo}%
  \providecommand \translation [1]{[#1]}%
  \providecommand \BibitemOpen [0]{}%
  \providecommand \bibitemStop [0]{}%
  \providecommand \bibitemNoStop [0]{.\EOS\space}%
  \providecommand \EOS [0]{\spacefactor3000\relax}%
  \providecommand \BibitemShut  [1]{\csname bibitem#1\endcsname}%
  \let\auto@bib@innerbib\@empty
  \bibitem [{\citenamefont {Emma}\ \emph {et~al.}(2010)\citenamefont {Emma},
    \citenamefont {Akre}, \citenamefont {Arthur}, \citenamefont {Bionta},
    \citenamefont {Bostedt}, \citenamefont {Bozek}, \citenamefont {Brachmann},
    \citenamefont {Bucksbaum}, \citenamefont {Coffee}, \citenamefont {Decker},
    \citenamefont {Ding}, \citenamefont {Dowell}, \citenamefont {Edstrom},
    \citenamefont {Fisher}, \citenamefont {Frisch}, \citenamefont {Gilevich},
    \citenamefont {Hastings}, \citenamefont {Hays}, \citenamefont {Hering},
    \citenamefont {Huang}, \citenamefont {Iverson}, \citenamefont {Loos},
    \citenamefont {Messerschmidt}, \citenamefont {Miahnahri}, \citenamefont
    {Moeller}, \citenamefont {Nuhn}, \citenamefont {Pile}, \citenamefont
    {Ratner}, \citenamefont {Rzepiela}, \citenamefont {Schultz}, \citenamefont
    {Smith}, \citenamefont {Stefan}, \citenamefont {Tompkins}, \citenamefont
    {Turner}, \citenamefont {Welch}, \citenamefont {White}, \citenamefont {Wu},
    \citenamefont {Yocky},\ and\ \citenamefont {Galayda}}]{emma2010first}%
    \BibitemOpen
    \bibfield  {author} {\bibinfo {author} {\bibfnamefont {P.}~\bibnamefont
    {Emma}}, \bibinfo {author} {\bibfnamefont {R.}~\bibnamefont {Akre}}, \bibinfo
    {author} {\bibfnamefont {J.}~\bibnamefont {Arthur}}, \bibinfo {author}
    {\bibfnamefont {R.}~\bibnamefont {Bionta}}, \bibinfo {author} {\bibfnamefont
    {C.}~\bibnamefont {Bostedt}}, \bibinfo {author} {\bibfnamefont
    {J.}~\bibnamefont {Bozek}}, \bibinfo {author} {\bibfnamefont
    {A.}~\bibnamefont {Brachmann}}, \bibinfo {author} {\bibfnamefont
    {P.}~\bibnamefont {Bucksbaum}}, \bibinfo {author} {\bibfnamefont
    {R.}~\bibnamefont {Coffee}}, \bibinfo {author} {\bibfnamefont {F.~J.}\
    \bibnamefont {Decker}}, \bibinfo {author} {\bibfnamefont {Y.}~\bibnamefont
    {Ding}}, \bibinfo {author} {\bibfnamefont {D.}~\bibnamefont {Dowell}},
    \bibinfo {author} {\bibfnamefont {S.}~\bibnamefont {Edstrom}}, \bibinfo
    {author} {\bibfnamefont {A.}~\bibnamefont {Fisher}}, \bibinfo {author}
    {\bibfnamefont {J.}~\bibnamefont {Frisch}}, \bibinfo {author} {\bibfnamefont
    {S.}~\bibnamefont {Gilevich}}, \bibinfo {author} {\bibfnamefont
    {J.}~\bibnamefont {Hastings}}, \bibinfo {author} {\bibfnamefont
    {G.}~\bibnamefont {Hays}}, \bibinfo {author} {\bibfnamefont {P.}~\bibnamefont
    {Hering}}, \bibinfo {author} {\bibfnamefont {Z.}~\bibnamefont {Huang}},
    \bibinfo {author} {\bibfnamefont {R.}~\bibnamefont {Iverson}}, \bibinfo
    {author} {\bibfnamefont {H.}~\bibnamefont {Loos}}, \bibinfo {author}
    {\bibfnamefont {M.}~\bibnamefont {Messerschmidt}}, \bibinfo {author}
    {\bibfnamefont {A.}~\bibnamefont {Miahnahri}}, \bibinfo {author}
    {\bibfnamefont {S.}~\bibnamefont {Moeller}}, \bibinfo {author} {\bibfnamefont
    {H.~D.}\ \bibnamefont {Nuhn}}, \bibinfo {author} {\bibfnamefont
    {G.}~\bibnamefont {Pile}}, \bibinfo {author} {\bibfnamefont {D.}~\bibnamefont
    {Ratner}}, \bibinfo {author} {\bibfnamefont {J.}~\bibnamefont {Rzepiela}},
    \bibinfo {author} {\bibfnamefont {D.}~\bibnamefont {Schultz}}, \bibinfo
    {author} {\bibfnamefont {T.}~\bibnamefont {Smith}}, \bibinfo {author}
    {\bibfnamefont {P.}~\bibnamefont {Stefan}}, \bibinfo {author} {\bibfnamefont
    {H.}~\bibnamefont {Tompkins}}, \bibinfo {author} {\bibfnamefont
    {J.}~\bibnamefont {Turner}}, \bibinfo {author} {\bibfnamefont
    {J.}~\bibnamefont {Welch}}, \bibinfo {author} {\bibfnamefont
    {W.}~\bibnamefont {White}}, \bibinfo {author} {\bibfnamefont
    {J.}~\bibnamefont {Wu}}, \bibinfo {author} {\bibfnamefont {G.}~\bibnamefont
    {Yocky}}, \ and\ \bibinfo {author} {\bibfnamefont {J.}~\bibnamefont
    {Galayda}},\ }\href {\doibase 10.1038/nphoton.2010.176} {\bibfield  {journal}
    {\bibinfo  {journal} {Nat. Photonics}\ }\textbf {\bibinfo {volume} {4}},\
    \bibinfo {pages} {641} (\bibinfo {year} {2010})}\BibitemShut {NoStop}%
  \bibitem [{\citenamefont {Ishikawa}\ \emph {et~al.}(2012)\citenamefont
    {Ishikawa}, \citenamefont {Aoyagi}, \citenamefont {Asaka}, \citenamefont
    {Asano}, \citenamefont {Azumi}, \citenamefont {Bizen}, \citenamefont {Ego},
    \citenamefont {Fukami}, \citenamefont {Fukui}, \citenamefont {Furukawa},\
    and\ \citenamefont {Others}}]{ishikawa2012compact}%
    \BibitemOpen
    \bibfield  {author} {\bibinfo {author} {\bibfnamefont {T.}~\bibnamefont
    {Ishikawa}}, \bibinfo {author} {\bibfnamefont {H.}~\bibnamefont {Aoyagi}},
    \bibinfo {author} {\bibfnamefont {T.}~\bibnamefont {Asaka}}, \bibinfo
    {author} {\bibfnamefont {Y.}~\bibnamefont {Asano}}, \bibinfo {author}
    {\bibfnamefont {N.}~\bibnamefont {Azumi}}, \bibinfo {author} {\bibfnamefont
    {T.}~\bibnamefont {Bizen}}, \bibinfo {author} {\bibfnamefont
    {H.}~\bibnamefont {Ego}}, \bibinfo {author} {\bibfnamefont {K.}~\bibnamefont
    {Fukami}}, \bibinfo {author} {\bibfnamefont {T.}~\bibnamefont {Fukui}},
    \bibinfo {author} {\bibfnamefont {Y.}~\bibnamefont {Furukawa}}, \ and\
    \bibinfo {author} {\bibnamefont {Others}},\ }\href@noop {} {\bibfield
    {journal} {\bibinfo  {journal} {Nat. Photonics}\ }\textbf {\bibinfo {volume}
    {6}},\ \bibinfo {pages} {540} (\bibinfo {year} {2012})}\BibitemShut {NoStop}%
  \bibitem [{\citenamefont {Ackermann}\ \emph {et~al.}(2007)\citenamefont
    {Ackermann}, \citenamefont {Asova}, \citenamefont {Ayvazyan}, \citenamefont
    {Azima}, \citenamefont {Baboi}, \citenamefont {B{\"{a}}hr}, \citenamefont
    {Balandin}, \citenamefont {Beutner}, \citenamefont {Brandt}, \citenamefont
    {Bolzmann}, \citenamefont {Brinkmann}, \citenamefont {Brovko}, \citenamefont
    {Castellano}, \citenamefont {Castro}, \citenamefont {Catani}, \citenamefont
    {Chiadroni}, \citenamefont {Choroba}, \citenamefont {Cianchi}, \citenamefont
    {Costello}, \citenamefont {Cubaynes}, \citenamefont {Dardis}, \citenamefont
    {Decking}, \citenamefont {Delsim-Hashemi}, \citenamefont {Delserieys},
    \citenamefont {{Di Pirro}}, \citenamefont {Dohlus}, \citenamefont
    {D{\"{u}}sterer}, \citenamefont {Eckhardt}, \citenamefont {Edwards},
    \citenamefont {Faatz}, \citenamefont {Feldhaus}, \citenamefont
    {Fl{\"{o}}ttmann}, \citenamefont {Frisch}, \citenamefont {Fr{\"{o}}hlich},
    \citenamefont {Garvey}, \citenamefont {Gensch}, \citenamefont {Gerth},
    \citenamefont {G{\"{o}}rler}, \citenamefont {Golubeva}, \citenamefont
    {Grabosch}, \citenamefont {Grecki}, \citenamefont {Grimm}, \citenamefont
    {Hacker}, \citenamefont {Hahn}, \citenamefont {Han}, \citenamefont
    {Honkavaara}, \citenamefont {Hott}, \citenamefont {H{\"{u}}ning},
    \citenamefont {Ivanisenko}, \citenamefont {Jaeschke}, \citenamefont
    {Jalmuzna}, \citenamefont {Jezynski}, \citenamefont {Kammering},
    \citenamefont {Katalev}, \citenamefont {Kavanagh}, \citenamefont {Kennedy},
    \citenamefont {Khodyachykh}, \citenamefont {Klose}, \citenamefont
    {Kocharyan}, \citenamefont {K{\"{o}}rfer}, \citenamefont {Kollewe},
    \citenamefont {Koprek}, \citenamefont {Korepanov}, \citenamefont {Kostin},
    \citenamefont {Krassilnikov}, \citenamefont {Kube}, \citenamefont {Kuhlmann},
    \citenamefont {Lewis}, \citenamefont {Lilje}, \citenamefont {Limberg},
    \citenamefont {Lipka}, \citenamefont {L{\"{o}}hl}, \citenamefont {Luna},
    \citenamefont {Luong}, \citenamefont {Martins}, \citenamefont {Meyer},
    \citenamefont {Michelato}, \citenamefont {Miltchev}, \citenamefont
    {M{\"{o}}ller}, \citenamefont {Monaco}, \citenamefont {M{\"{u}}ller},
    \citenamefont {Napieralski}, \citenamefont {Napoly}, \citenamefont
    {Nicolosi}, \citenamefont {N{\"{o}}lle}, \citenamefont {Nũez}, \citenamefont
    {Oppelt}, \citenamefont {Pagani}, \citenamefont {Paparella}, \citenamefont
    {Pchalek}, \citenamefont {Pedregosa-Gutierrez}, \citenamefont {Petersen},
    \citenamefont {Petrosyan}, \citenamefont {Petrosyan}, \citenamefont
    {Petrosyan}, \citenamefont {Pfl{\"{u}}ger}, \citenamefont {Pl{\"{o}}njes},
    \citenamefont {Poletto}, \citenamefont {Pozniak}, \citenamefont {Prat},
    \citenamefont {Proch}, \citenamefont {Pucyk}, \citenamefont {Radcliffe},
    \citenamefont {Redlin}, \citenamefont {Rehlich}, \citenamefont {Richter},
    \citenamefont {Roehrs}, \citenamefont {Roensch}, \citenamefont {Romaniuk},
    \citenamefont {Ross}, \citenamefont {Rossbach}, \citenamefont {Rybnikov},
    \citenamefont {Sachwitz}, \citenamefont {Saldin}, \citenamefont {Sandner},
    \citenamefont {Schlarb}, \citenamefont {Schmidt}, \citenamefont {Schmitz},
    \citenamefont {Schm{\"{u}}ser}, \citenamefont {Schneider}, \citenamefont
    {Schneidmiller}, \citenamefont {Schnepp}, \citenamefont {Schreiber},
    \citenamefont {Seidel}, \citenamefont {Sertore}, \citenamefont {Shabunov},
    \citenamefont {Simon}, \citenamefont {Simrock}, \citenamefont {Sombrowski},
    \citenamefont {Sorokin}, \citenamefont {Spanknebel}, \citenamefont
    {Spesyvtsev}, \citenamefont {Staykov}, \citenamefont {Steffen}, \citenamefont
    {Stephan}, \citenamefont {Stulle}, \citenamefont {Thom}, \citenamefont
    {Tiedtke}, \citenamefont {Tischer}, \citenamefont {Toleikis}, \citenamefont
    {Treusch}, \citenamefont {Trines}, \citenamefont {Tsakov}, \citenamefont
    {Vogel}, \citenamefont {Weiland}, \citenamefont {Weise}, \citenamefont
    {Wellh{\"{o}}fer}, \citenamefont {Wendt}, \citenamefont {Will}, \citenamefont
    {Winter}, \citenamefont {Wittenburg}, \citenamefont {Wurth}, \citenamefont
    {Yeates}, \citenamefont {Yurkov}, \citenamefont {Zagorodnov},\ and\
    \citenamefont {Zapfe}}]{Ackermann.2007}%
    \BibitemOpen
    \bibfield  {author} {\bibinfo {author} {\bibfnamefont {W.}~\bibnamefont
    {Ackermann}}, \bibinfo {author} {\bibfnamefont {G.}~\bibnamefont {Asova}},
    \bibinfo {author} {\bibfnamefont {V.}~\bibnamefont {Ayvazyan}}, \bibinfo
    {author} {\bibfnamefont {A.}~\bibnamefont {Azima}}, \bibinfo {author}
    {\bibfnamefont {N.}~\bibnamefont {Baboi}}, \bibinfo {author} {\bibfnamefont
    {J.}~\bibnamefont {B{\"{a}}hr}}, \bibinfo {author} {\bibfnamefont
    {V.}~\bibnamefont {Balandin}}, \bibinfo {author} {\bibfnamefont
    {B.}~\bibnamefont {Beutner}}, \bibinfo {author} {\bibfnamefont
    {A.}~\bibnamefont {Brandt}}, \bibinfo {author} {\bibfnamefont
    {A.}~\bibnamefont {Bolzmann}}, \bibinfo {author} {\bibfnamefont
    {R.}~\bibnamefont {Brinkmann}}, \bibinfo {author} {\bibfnamefont {O.~I.}\
    \bibnamefont {Brovko}}, \bibinfo {author} {\bibfnamefont {M.}~\bibnamefont
    {Castellano}}, \bibinfo {author} {\bibfnamefont {P.}~\bibnamefont {Castro}},
    \bibinfo {author} {\bibfnamefont {L.}~\bibnamefont {Catani}}, \bibinfo
    {author} {\bibfnamefont {E.}~\bibnamefont {Chiadroni}}, \bibinfo {author}
    {\bibfnamefont {S.}~\bibnamefont {Choroba}}, \bibinfo {author} {\bibfnamefont
    {A.}~\bibnamefont {Cianchi}}, \bibinfo {author} {\bibfnamefont {J.~T.}\
    \bibnamefont {Costello}}, \bibinfo {author} {\bibfnamefont {D.}~\bibnamefont
    {Cubaynes}}, \bibinfo {author} {\bibfnamefont {J.}~\bibnamefont {Dardis}},
    \bibinfo {author} {\bibfnamefont {W.}~\bibnamefont {Decking}}, \bibinfo
    {author} {\bibfnamefont {H.}~\bibnamefont {Delsim-Hashemi}}, \bibinfo
    {author} {\bibfnamefont {A.}~\bibnamefont {Delserieys}}, \bibinfo {author}
    {\bibfnamefont {G.}~\bibnamefont {{Di Pirro}}}, \bibinfo {author}
    {\bibfnamefont {M.}~\bibnamefont {Dohlus}}, \bibinfo {author} {\bibfnamefont
    {S.}~\bibnamefont {D{\"{u}}sterer}}, \bibinfo {author} {\bibfnamefont
    {A.}~\bibnamefont {Eckhardt}}, \bibinfo {author} {\bibfnamefont {H.~T.}\
    \bibnamefont {Edwards}}, \bibinfo {author} {\bibfnamefont {B.}~\bibnamefont
    {Faatz}}, \bibinfo {author} {\bibfnamefont {J.}~\bibnamefont {Feldhaus}},
    \bibinfo {author} {\bibfnamefont {K.}~\bibnamefont {Fl{\"{o}}ttmann}},
    \bibinfo {author} {\bibfnamefont {J.}~\bibnamefont {Frisch}}, \bibinfo
    {author} {\bibfnamefont {L.}~\bibnamefont {Fr{\"{o}}hlich}}, \bibinfo
    {author} {\bibfnamefont {T.}~\bibnamefont {Garvey}}, \bibinfo {author}
    {\bibfnamefont {U.}~\bibnamefont {Gensch}}, \bibinfo {author} {\bibfnamefont
    {C.}~\bibnamefont {Gerth}}, \bibinfo {author} {\bibfnamefont
    {M.}~\bibnamefont {G{\"{o}}rler}}, \bibinfo {author} {\bibfnamefont
    {N.}~\bibnamefont {Golubeva}}, \bibinfo {author} {\bibfnamefont {H.~J.}\
    \bibnamefont {Grabosch}}, \bibinfo {author} {\bibfnamefont {M.}~\bibnamefont
    {Grecki}}, \bibinfo {author} {\bibfnamefont {O.}~\bibnamefont {Grimm}},
    \bibinfo {author} {\bibfnamefont {K.}~\bibnamefont {Hacker}}, \bibinfo
    {author} {\bibfnamefont {U.}~\bibnamefont {Hahn}}, \bibinfo {author}
    {\bibfnamefont {J.~H.}\ \bibnamefont {Han}}, \bibinfo {author} {\bibfnamefont
    {K.}~\bibnamefont {Honkavaara}}, \bibinfo {author} {\bibfnamefont
    {T.}~\bibnamefont {Hott}}, \bibinfo {author} {\bibfnamefont {M.}~\bibnamefont
    {H{\"{u}}ning}}, \bibinfo {author} {\bibfnamefont {Y.}~\bibnamefont
    {Ivanisenko}}, \bibinfo {author} {\bibfnamefont {E.}~\bibnamefont
    {Jaeschke}}, \bibinfo {author} {\bibfnamefont {W.}~\bibnamefont {Jalmuzna}},
    \bibinfo {author} {\bibfnamefont {T.}~\bibnamefont {Jezynski}}, \bibinfo
    {author} {\bibfnamefont {R.}~\bibnamefont {Kammering}}, \bibinfo {author}
    {\bibfnamefont {V.}~\bibnamefont {Katalev}}, \bibinfo {author} {\bibfnamefont
    {K.}~\bibnamefont {Kavanagh}}, \bibinfo {author} {\bibfnamefont {E.~T.}\
    \bibnamefont {Kennedy}}, \bibinfo {author} {\bibfnamefont {S.}~\bibnamefont
    {Khodyachykh}}, \bibinfo {author} {\bibfnamefont {K.}~\bibnamefont {Klose}},
    \bibinfo {author} {\bibfnamefont {V.}~\bibnamefont {Kocharyan}}, \bibinfo
    {author} {\bibfnamefont {M.}~\bibnamefont {K{\"{o}}rfer}}, \bibinfo {author}
    {\bibfnamefont {M.}~\bibnamefont {Kollewe}}, \bibinfo {author} {\bibfnamefont
    {W.}~\bibnamefont {Koprek}}, \bibinfo {author} {\bibfnamefont
    {S.}~\bibnamefont {Korepanov}}, \bibinfo {author} {\bibfnamefont
    {D.}~\bibnamefont {Kostin}}, \bibinfo {author} {\bibfnamefont
    {M.}~\bibnamefont {Krassilnikov}}, \bibinfo {author} {\bibfnamefont
    {G.}~\bibnamefont {Kube}}, \bibinfo {author} {\bibfnamefont {M.}~\bibnamefont
    {Kuhlmann}}, \bibinfo {author} {\bibfnamefont {C.~L.}\ \bibnamefont {Lewis}},
    \bibinfo {author} {\bibfnamefont {L.}~\bibnamefont {Lilje}}, \bibinfo
    {author} {\bibfnamefont {T.}~\bibnamefont {Limberg}}, \bibinfo {author}
    {\bibfnamefont {D.}~\bibnamefont {Lipka}}, \bibinfo {author} {\bibfnamefont
    {F.}~\bibnamefont {L{\"{o}}hl}}, \bibinfo {author} {\bibfnamefont
    {H.}~\bibnamefont {Luna}}, \bibinfo {author} {\bibfnamefont {M.}~\bibnamefont
    {Luong}}, \bibinfo {author} {\bibfnamefont {M.}~\bibnamefont {Martins}},
    \bibinfo {author} {\bibfnamefont {M.}~\bibnamefont {Meyer}}, \bibinfo
    {author} {\bibfnamefont {P.}~\bibnamefont {Michelato}}, \bibinfo {author}
    {\bibfnamefont {V.}~\bibnamefont {Miltchev}}, \bibinfo {author}
    {\bibfnamefont {W.~D.}\ \bibnamefont {M{\"{o}}ller}}, \bibinfo {author}
    {\bibfnamefont {L.}~\bibnamefont {Monaco}}, \bibinfo {author} {\bibfnamefont
    {W.~F.}\ \bibnamefont {M{\"{u}}ller}}, \bibinfo {author} {\bibfnamefont
    {O.}~\bibnamefont {Napieralski}}, \bibinfo {author} {\bibfnamefont
    {O.}~\bibnamefont {Napoly}}, \bibinfo {author} {\bibfnamefont
    {P.}~\bibnamefont {Nicolosi}}, \bibinfo {author} {\bibfnamefont
    {D.}~\bibnamefont {N{\"{o}}lle}}, \bibinfo {author} {\bibfnamefont
    {T.}~\bibnamefont {Nũez}}, \bibinfo {author} {\bibfnamefont
    {A.}~\bibnamefont {Oppelt}}, \bibinfo {author} {\bibfnamefont
    {C.}~\bibnamefont {Pagani}}, \bibinfo {author} {\bibfnamefont
    {R.}~\bibnamefont {Paparella}}, \bibinfo {author} {\bibfnamefont
    {N.}~\bibnamefont {Pchalek}}, \bibinfo {author} {\bibfnamefont
    {J.}~\bibnamefont {Pedregosa-Gutierrez}}, \bibinfo {author} {\bibfnamefont
    {B.}~\bibnamefont {Petersen}}, \bibinfo {author} {\bibfnamefont
    {B.}~\bibnamefont {Petrosyan}}, \bibinfo {author} {\bibfnamefont
    {G.}~\bibnamefont {Petrosyan}}, \bibinfo {author} {\bibfnamefont
    {L.}~\bibnamefont {Petrosyan}}, \bibinfo {author} {\bibfnamefont
    {J.}~\bibnamefont {Pfl{\"{u}}ger}}, \bibinfo {author} {\bibfnamefont
    {E.}~\bibnamefont {Pl{\"{o}}njes}}, \bibinfo {author} {\bibfnamefont
    {L.}~\bibnamefont {Poletto}}, \bibinfo {author} {\bibfnamefont
    {K.}~\bibnamefont {Pozniak}}, \bibinfo {author} {\bibfnamefont
    {E.}~\bibnamefont {Prat}}, \bibinfo {author} {\bibfnamefont {D.}~\bibnamefont
    {Proch}}, \bibinfo {author} {\bibfnamefont {P.}~\bibnamefont {Pucyk}},
    \bibinfo {author} {\bibfnamefont {P.}~\bibnamefont {Radcliffe}}, \bibinfo
    {author} {\bibfnamefont {H.}~\bibnamefont {Redlin}}, \bibinfo {author}
    {\bibfnamefont {K.}~\bibnamefont {Rehlich}}, \bibinfo {author} {\bibfnamefont
    {M.}~\bibnamefont {Richter}}, \bibinfo {author} {\bibfnamefont
    {M.}~\bibnamefont {Roehrs}}, \bibinfo {author} {\bibfnamefont
    {J.}~\bibnamefont {Roensch}}, \bibinfo {author} {\bibfnamefont
    {R.}~\bibnamefont {Romaniuk}}, \bibinfo {author} {\bibfnamefont
    {M.}~\bibnamefont {Ross}}, \bibinfo {author} {\bibfnamefont {J.}~\bibnamefont
    {Rossbach}}, \bibinfo {author} {\bibfnamefont {V.}~\bibnamefont {Rybnikov}},
    \bibinfo {author} {\bibfnamefont {M.}~\bibnamefont {Sachwitz}}, \bibinfo
    {author} {\bibfnamefont {E.~L.}\ \bibnamefont {Saldin}}, \bibinfo {author}
    {\bibfnamefont {W.}~\bibnamefont {Sandner}}, \bibinfo {author} {\bibfnamefont
    {H.}~\bibnamefont {Schlarb}}, \bibinfo {author} {\bibfnamefont
    {B.}~\bibnamefont {Schmidt}}, \bibinfo {author} {\bibfnamefont
    {M.}~\bibnamefont {Schmitz}}, \bibinfo {author} {\bibfnamefont
    {P.}~\bibnamefont {Schm{\"{u}}ser}}, \bibinfo {author} {\bibfnamefont
    {J.~R.}\ \bibnamefont {Schneider}}, \bibinfo {author} {\bibfnamefont {E.~A.}\
    \bibnamefont {Schneidmiller}}, \bibinfo {author} {\bibfnamefont
    {S.}~\bibnamefont {Schnepp}}, \bibinfo {author} {\bibfnamefont
    {S.}~\bibnamefont {Schreiber}}, \bibinfo {author} {\bibfnamefont
    {M.}~\bibnamefont {Seidel}}, \bibinfo {author} {\bibfnamefont
    {D.}~\bibnamefont {Sertore}}, \bibinfo {author} {\bibfnamefont {A.~V.}\
    \bibnamefont {Shabunov}}, \bibinfo {author} {\bibfnamefont {C.}~\bibnamefont
    {Simon}}, \bibinfo {author} {\bibfnamefont {S.}~\bibnamefont {Simrock}},
    \bibinfo {author} {\bibfnamefont {E.}~\bibnamefont {Sombrowski}}, \bibinfo
    {author} {\bibfnamefont {A.~A.}\ \bibnamefont {Sorokin}}, \bibinfo {author}
    {\bibfnamefont {P.}~\bibnamefont {Spanknebel}}, \bibinfo {author}
    {\bibfnamefont {R.}~\bibnamefont {Spesyvtsev}}, \bibinfo {author}
    {\bibfnamefont {L.}~\bibnamefont {Staykov}}, \bibinfo {author} {\bibfnamefont
    {B.}~\bibnamefont {Steffen}}, \bibinfo {author} {\bibfnamefont
    {F.}~\bibnamefont {Stephan}}, \bibinfo {author} {\bibfnamefont
    {F.}~\bibnamefont {Stulle}}, \bibinfo {author} {\bibfnamefont
    {H.}~\bibnamefont {Thom}}, \bibinfo {author} {\bibfnamefont {K.}~\bibnamefont
    {Tiedtke}}, \bibinfo {author} {\bibfnamefont {M.}~\bibnamefont {Tischer}},
    \bibinfo {author} {\bibfnamefont {S.}~\bibnamefont {Toleikis}}, \bibinfo
    {author} {\bibfnamefont {R.}~\bibnamefont {Treusch}}, \bibinfo {author}
    {\bibfnamefont {D.}~\bibnamefont {Trines}}, \bibinfo {author} {\bibfnamefont
    {I.}~\bibnamefont {Tsakov}}, \bibinfo {author} {\bibfnamefont
    {E.}~\bibnamefont {Vogel}}, \bibinfo {author} {\bibfnamefont
    {T.}~\bibnamefont {Weiland}}, \bibinfo {author} {\bibfnamefont
    {H.}~\bibnamefont {Weise}}, \bibinfo {author} {\bibfnamefont
    {M.}~\bibnamefont {Wellh{\"{o}}fer}}, \bibinfo {author} {\bibfnamefont
    {M.}~\bibnamefont {Wendt}}, \bibinfo {author} {\bibfnamefont
    {I.}~\bibnamefont {Will}}, \bibinfo {author} {\bibfnamefont {A.}~\bibnamefont
    {Winter}}, \bibinfo {author} {\bibfnamefont {K.}~\bibnamefont {Wittenburg}},
    \bibinfo {author} {\bibfnamefont {W.}~\bibnamefont {Wurth}}, \bibinfo
    {author} {\bibfnamefont {P.}~\bibnamefont {Yeates}}, \bibinfo {author}
    {\bibfnamefont {M.~V.}\ \bibnamefont {Yurkov}}, \bibinfo {author}
    {\bibfnamefont {I.}~\bibnamefont {Zagorodnov}}, \ and\ \bibinfo {author}
    {\bibfnamefont {K.}~\bibnamefont {Zapfe}},\ }\href {\doibase
    10.1038/nphoton.2007.76} {\bibfield  {journal} {\bibinfo  {journal} {Nat.
    Photonics}\ }\textbf {\bibinfo {volume} {1}},\ \bibinfo {pages} {336}
    (\bibinfo {year} {2007})}\BibitemShut {NoStop}%
  \bibitem [{\citenamefont {Kang}\ \emph {et~al.}(2017)\citenamefont {Kang},
    \citenamefont {Min}, \citenamefont {Heo}, \citenamefont {Kim}, \citenamefont
    {Yang}, \citenamefont {Kim}, \citenamefont {Nam}, \citenamefont {Baek},
    \citenamefont {Choi}, \citenamefont {Mun}, \citenamefont {Park},
    \citenamefont {Suh}, \citenamefont {Shin}, \citenamefont {Hu}, \citenamefont
    {Hong}, \citenamefont {Jung}, \citenamefont {Kim}, \citenamefont {Kim},
    \citenamefont {Na}, \citenamefont {Park}, \citenamefont {Park}, \citenamefont
    {Han}, \citenamefont {Jung}, \citenamefont {Jeong}, \citenamefont {Lee},
    \citenamefont {Lee}, \citenamefont {Lee}, \citenamefont {Lee}, \citenamefont
    {Oh}, \citenamefont {Suh}, \citenamefont {Parc}, \citenamefont {Park},
    \citenamefont {Kim}, \citenamefont {Jung}, \citenamefont {Kim}, \citenamefont
    {Lee}, \citenamefont {Lee}, \citenamefont {Sung}, \citenamefont {Mok},
    \citenamefont {Yang}, \citenamefont {Lee}, \citenamefont {Shin},
    \citenamefont {Kim}, \citenamefont {Kim}, \citenamefont {Lee}, \citenamefont
    {Park}, \citenamefont {Kim}, \citenamefont {Park}, \citenamefont {Eom},
    \citenamefont {Rah}, \citenamefont {Kim}, \citenamefont {Nam}, \citenamefont
    {Park}, \citenamefont {Park}, \citenamefont {Kim}, \citenamefont {Kwon},
    \citenamefont {Park}, \citenamefont {Kim}, \citenamefont {Hyun},
    \citenamefont {Kim}, \citenamefont {Kim}, \citenamefont {Hwang},
    \citenamefont {Kim}, \citenamefont {Lim}, \citenamefont {Yu}, \citenamefont
    {Kim}, \citenamefont {Kang}, \citenamefont {Kim}, \citenamefont {Kim},
    \citenamefont {Lee}, \citenamefont {Lee}, \citenamefont {Park}, \citenamefont
    {Koo}, \citenamefont {Kim},\ and\ \citenamefont {Ko}}]{Kang2017}%
    \BibitemOpen
    \bibfield  {author} {\bibinfo {author} {\bibfnamefont {H.~S.}\ \bibnamefont
    {Kang}}, \bibinfo {author} {\bibfnamefont {C.~K.}\ \bibnamefont {Min}},
    \bibinfo {author} {\bibfnamefont {H.}~\bibnamefont {Heo}}, \bibinfo {author}
    {\bibfnamefont {C.}~\bibnamefont {Kim}}, \bibinfo {author} {\bibfnamefont
    {H.}~\bibnamefont {Yang}}, \bibinfo {author} {\bibfnamefont {G.}~\bibnamefont
    {Kim}}, \bibinfo {author} {\bibfnamefont {I.}~\bibnamefont {Nam}}, \bibinfo
    {author} {\bibfnamefont {S.~Y.}\ \bibnamefont {Baek}}, \bibinfo {author}
    {\bibfnamefont {H.~J.}\ \bibnamefont {Choi}}, \bibinfo {author}
    {\bibfnamefont {G.}~\bibnamefont {Mun}}, \bibinfo {author} {\bibfnamefont
    {B.~R.}\ \bibnamefont {Park}}, \bibinfo {author} {\bibfnamefont {Y.~J.}\
    \bibnamefont {Suh}}, \bibinfo {author} {\bibfnamefont {D.~C.}\ \bibnamefont
    {Shin}}, \bibinfo {author} {\bibfnamefont {J.}~\bibnamefont {Hu}}, \bibinfo
    {author} {\bibfnamefont {J.}~\bibnamefont {Hong}}, \bibinfo {author}
    {\bibfnamefont {S.}~\bibnamefont {Jung}}, \bibinfo {author} {\bibfnamefont
    {S.~H.}\ \bibnamefont {Kim}}, \bibinfo {author} {\bibfnamefont {K.~H.}\
    \bibnamefont {Kim}}, \bibinfo {author} {\bibfnamefont {D.}~\bibnamefont
    {Na}}, \bibinfo {author} {\bibfnamefont {S.~S.}\ \bibnamefont {Park}},
    \bibinfo {author} {\bibfnamefont {Y.~J.}\ \bibnamefont {Park}}, \bibinfo
    {author} {\bibfnamefont {J.~H.}\ \bibnamefont {Han}}, \bibinfo {author}
    {\bibfnamefont {Y.~G.}\ \bibnamefont {Jung}}, \bibinfo {author}
    {\bibfnamefont {S.~H.}\ \bibnamefont {Jeong}}, \bibinfo {author}
    {\bibfnamefont {H.~G.}\ \bibnamefont {Lee}}, \bibinfo {author} {\bibfnamefont
    {S.}~\bibnamefont {Lee}}, \bibinfo {author} {\bibfnamefont {S.}~\bibnamefont
    {Lee}}, \bibinfo {author} {\bibfnamefont {W.~W.}\ \bibnamefont {Lee}},
    \bibinfo {author} {\bibfnamefont {B.}~\bibnamefont {Oh}}, \bibinfo {author}
    {\bibfnamefont {H.~S.}\ \bibnamefont {Suh}}, \bibinfo {author} {\bibfnamefont
    {Y.~W.}\ \bibnamefont {Parc}}, \bibinfo {author} {\bibfnamefont {S.~J.}\
    \bibnamefont {Park}}, \bibinfo {author} {\bibfnamefont {M.~H.}\ \bibnamefont
    {Kim}}, \bibinfo {author} {\bibfnamefont {N.~S.}\ \bibnamefont {Jung}},
    \bibinfo {author} {\bibfnamefont {Y.~C.}\ \bibnamefont {Kim}}, \bibinfo
    {author} {\bibfnamefont {M.~S.}\ \bibnamefont {Lee}}, \bibinfo {author}
    {\bibfnamefont {B.~H.}\ \bibnamefont {Lee}}, \bibinfo {author} {\bibfnamefont
    {C.~W.}\ \bibnamefont {Sung}}, \bibinfo {author} {\bibfnamefont {I.~S.}\
    \bibnamefont {Mok}}, \bibinfo {author} {\bibfnamefont {J.~M.}\ \bibnamefont
    {Yang}}, \bibinfo {author} {\bibfnamefont {C.~S.}\ \bibnamefont {Lee}},
    \bibinfo {author} {\bibfnamefont {H.}~\bibnamefont {Shin}}, \bibinfo {author}
    {\bibfnamefont {J.~H.}\ \bibnamefont {Kim}}, \bibinfo {author} {\bibfnamefont
    {Y.}~\bibnamefont {Kim}}, \bibinfo {author} {\bibfnamefont {J.~H.}\
    \bibnamefont {Lee}}, \bibinfo {author} {\bibfnamefont {S.~Y.}\ \bibnamefont
    {Park}}, \bibinfo {author} {\bibfnamefont {J.}~\bibnamefont {Kim}}, \bibinfo
    {author} {\bibfnamefont {J.}~\bibnamefont {Park}}, \bibinfo {author}
    {\bibfnamefont {I.}~\bibnamefont {Eom}}, \bibinfo {author} {\bibfnamefont
    {S.}~\bibnamefont {Rah}}, \bibinfo {author} {\bibfnamefont {S.}~\bibnamefont
    {Kim}}, \bibinfo {author} {\bibfnamefont {K.~H.}\ \bibnamefont {Nam}},
    \bibinfo {author} {\bibfnamefont {J.}~\bibnamefont {Park}}, \bibinfo {author}
    {\bibfnamefont {J.}~\bibnamefont {Park}}, \bibinfo {author} {\bibfnamefont
    {S.}~\bibnamefont {Kim}}, \bibinfo {author} {\bibfnamefont {S.}~\bibnamefont
    {Kwon}}, \bibinfo {author} {\bibfnamefont {S.~H.}\ \bibnamefont {Park}},
    \bibinfo {author} {\bibfnamefont {K.~S.}\ \bibnamefont {Kim}}, \bibinfo
    {author} {\bibfnamefont {H.}~\bibnamefont {Hyun}}, \bibinfo {author}
    {\bibfnamefont {S.~N.}\ \bibnamefont {Kim}}, \bibinfo {author} {\bibfnamefont
    {S.}~\bibnamefont {Kim}}, \bibinfo {author} {\bibfnamefont {S.~M.}\
    \bibnamefont {Hwang}}, \bibinfo {author} {\bibfnamefont {M.~J.}\ \bibnamefont
    {Kim}}, \bibinfo {author} {\bibfnamefont {C.~Y.}\ \bibnamefont {Lim}},
    \bibinfo {author} {\bibfnamefont {C.~J.}\ \bibnamefont {Yu}}, \bibinfo
    {author} {\bibfnamefont {B.~S.}\ \bibnamefont {Kim}}, \bibinfo {author}
    {\bibfnamefont {T.~H.}\ \bibnamefont {Kang}}, \bibinfo {author}
    {\bibfnamefont {K.~W.}\ \bibnamefont {Kim}}, \bibinfo {author} {\bibfnamefont
    {S.~H.}\ \bibnamefont {Kim}}, \bibinfo {author} {\bibfnamefont {H.~S.}\
    \bibnamefont {Lee}}, \bibinfo {author} {\bibfnamefont {H.~S.}\ \bibnamefont
    {Lee}}, \bibinfo {author} {\bibfnamefont {K.~H.}\ \bibnamefont {Park}},
    \bibinfo {author} {\bibfnamefont {T.~Y.}\ \bibnamefont {Koo}}, \bibinfo
    {author} {\bibfnamefont {D.~E.}\ \bibnamefont {Kim}}, \ and\ \bibinfo
    {author} {\bibfnamefont {I.~S.}\ \bibnamefont {Ko}},\ }\href {\doibase
    10.1038/s41566-017-0029-8} {\bibfield  {journal} {\bibinfo  {journal} {Nat.
    Photonics}\ }\textbf {\bibinfo {volume} {11}},\ \bibinfo {pages} {708}
    (\bibinfo {year} {2017})}\BibitemShut {NoStop}%
  \bibitem [{\citenamefont {Milne}\ \emph {et~al.}(2017)\citenamefont {Milne},
    \citenamefont {Schietinger}, \citenamefont {Aiba}, \citenamefont {Alarcon},
    \citenamefont {Alex}, \citenamefont {Anghel}, \citenamefont {Arsov},
    \citenamefont {Beard}, \citenamefont {Beaud}, \citenamefont {Bettoni},\ and\
    \citenamefont {Others}}]{milne2017swissfel}%
    \BibitemOpen
    \bibfield  {author} {\bibinfo {author} {\bibfnamefont {C.}~\bibnamefont
    {Milne}}, \bibinfo {author} {\bibfnamefont {T.}~\bibnamefont {Schietinger}},
    \bibinfo {author} {\bibfnamefont {M.}~\bibnamefont {Aiba}}, \bibinfo {author}
    {\bibfnamefont {A.}~\bibnamefont {Alarcon}}, \bibinfo {author} {\bibfnamefont
    {J.}~\bibnamefont {Alex}}, \bibinfo {author} {\bibfnamefont {A.}~\bibnamefont
    {Anghel}}, \bibinfo {author} {\bibfnamefont {V.}~\bibnamefont {Arsov}},
    \bibinfo {author} {\bibfnamefont {C.}~\bibnamefont {Beard}}, \bibinfo
    {author} {\bibfnamefont {P.}~\bibnamefont {Beaud}}, \bibinfo {author}
    {\bibfnamefont {S.}~\bibnamefont {Bettoni}}, \ and\ \bibinfo {author}
    {\bibnamefont {Others}},\ }\href@noop {} {\bibfield  {journal} {\bibinfo
    {journal} {Appl. Sci.}\ }\textbf {\bibinfo {volume} {7}},\ \bibinfo {pages}
    {720} (\bibinfo {year} {2017})}\BibitemShut {NoStop}%
  \bibitem [{\citenamefont {Amann}\ \emph {et~al.}(2012)\citenamefont {Amann},
    \citenamefont {Berg}, \citenamefont {Blank}, \citenamefont {Decker},
    \citenamefont {Ding}, \citenamefont {Emma}, \citenamefont {Feng},
    \citenamefont {Frisch}, \citenamefont {Fritz}, \citenamefont {Hastings},\
    and\ \citenamefont {Others}}]{amann2012demonstration}%
    \BibitemOpen
    \bibfield  {author} {\bibinfo {author} {\bibfnamefont {J.}~\bibnamefont
    {Amann}}, \bibinfo {author} {\bibfnamefont {W.}~\bibnamefont {Berg}},
    \bibinfo {author} {\bibfnamefont {V.}~\bibnamefont {Blank}}, \bibinfo
    {author} {\bibfnamefont {F.-J.}\ \bibnamefont {Decker}}, \bibinfo {author}
    {\bibfnamefont {Y.}~\bibnamefont {Ding}}, \bibinfo {author} {\bibfnamefont
    {P.}~\bibnamefont {Emma}}, \bibinfo {author} {\bibfnamefont {Y.}~\bibnamefont
    {Feng}}, \bibinfo {author} {\bibfnamefont {J.}~\bibnamefont {Frisch}},
    \bibinfo {author} {\bibfnamefont {D.}~\bibnamefont {Fritz}}, \bibinfo
    {author} {\bibfnamefont {J.}~\bibnamefont {Hastings}}, \ and\ \bibinfo
    {author} {\bibnamefont {Others}},\ }\href@noop {} {\bibfield  {journal}
    {\bibinfo  {journal} {Nat. Photonics}\ }\textbf {\bibinfo {volume} {6}},\
    \bibinfo {pages} {693} (\bibinfo {year} {2012})}\BibitemShut {NoStop}%
  \bibitem [{\citenamefont {Feng}\ and\ \citenamefont {Deng}(2018)}]{Feng.2018}%
    \BibitemOpen
    \bibfield  {author} {\bibinfo {author} {\bibfnamefont {C.}~\bibnamefont
    {Feng}}\ and\ \bibinfo {author} {\bibfnamefont {H.~X.}\ \bibnamefont
    {Deng}},\ }\href {\doibase 10.1007/s41365-018-0490-1} {\bibfield  {journal}
    {\bibinfo  {journal} {Nucl. Sci. Tech.}\ }\textbf {\bibinfo {volume} {29}},\
    \bibinfo {pages} {160} (\bibinfo {year} {2018})}\BibitemShut {NoStop}%
  \bibitem [{\citenamefont {Marinelli}\ \emph {et~al.}(2015)\citenamefont
    {Marinelli}, \citenamefont {Ratner}, \citenamefont {Lutman}, \citenamefont
    {Turner}, \citenamefont {Welch}, \citenamefont {Decker}, \citenamefont
    {Loos}, \citenamefont {Behrens}, \citenamefont {Gilevich}, \citenamefont
    {Miahnahri}, \citenamefont {Vetter}, \citenamefont {Maxwell}, \citenamefont
    {Ding}, \citenamefont {Coffee}, \citenamefont {Wakatsuki},\ and\
    \citenamefont {Huang}}]{Marinelli.2015}%
    \BibitemOpen
    \bibfield  {author} {\bibinfo {author} {\bibfnamefont {A.}~\bibnamefont
    {Marinelli}}, \bibinfo {author} {\bibfnamefont {D.}~\bibnamefont {Ratner}},
    \bibinfo {author} {\bibfnamefont {A.~A.}\ \bibnamefont {Lutman}}, \bibinfo
    {author} {\bibfnamefont {J.}~\bibnamefont {Turner}}, \bibinfo {author}
    {\bibfnamefont {J.}~\bibnamefont {Welch}}, \bibinfo {author} {\bibfnamefont
    {F.~J.}\ \bibnamefont {Decker}}, \bibinfo {author} {\bibfnamefont
    {H.}~\bibnamefont {Loos}}, \bibinfo {author} {\bibfnamefont {C.}~\bibnamefont
    {Behrens}}, \bibinfo {author} {\bibfnamefont {S.}~\bibnamefont {Gilevich}},
    \bibinfo {author} {\bibfnamefont {A.~A.}\ \bibnamefont {Miahnahri}}, \bibinfo
    {author} {\bibfnamefont {S.}~\bibnamefont {Vetter}}, \bibinfo {author}
    {\bibfnamefont {T.~J.}\ \bibnamefont {Maxwell}}, \bibinfo {author}
    {\bibfnamefont {Y.}~\bibnamefont {Ding}}, \bibinfo {author} {\bibfnamefont
    {R.}~\bibnamefont {Coffee}}, \bibinfo {author} {\bibfnamefont
    {S.}~\bibnamefont {Wakatsuki}}, \ and\ \bibinfo {author} {\bibfnamefont
    {Z.}~\bibnamefont {Huang}},\ }\href {\doibase 10.1038/ncomms7369} {\bibfield
    {journal} {\bibinfo  {journal} {Nat. Commun.}\ }\textbf {\bibinfo {volume}
    {6}},\ \bibinfo {pages} {6369} (\bibinfo {year} {2015})}\BibitemShut
    {NoStop}%
  \bibitem [{\citenamefont {Qiang}\ and\ \citenamefont
    {Wu}(2010)}]{qiang2011generation}%
    \BibitemOpen
    \bibfield  {author} {\bibinfo {author} {\bibfnamefont {J.}~\bibnamefont
    {Qiang}}\ and\ \bibinfo {author} {\bibfnamefont {J.}~\bibnamefont {Wu}},\
    }\href@noop {} {\bibfield  {journal} {\bibinfo  {journal} {FEL 2010 - 32nd
    Int. Free Electron Laser Conf.}\ }\textbf {\bibinfo {volume} {99}},\ \bibinfo
    {pages} {290} (\bibinfo {year} {2010})}\BibitemShut {NoStop}%
  \bibitem [{\citenamefont {Lutman}\ \emph
    {et~al.}(2016{\natexlab{a}})\citenamefont {Lutman}, \citenamefont {Maxwell},
    \citenamefont {MacArthur}, \citenamefont {Guetg}, \citenamefont {Berrah},
    \citenamefont {Coffee}, \citenamefont {Ding}, \citenamefont {Huang},
    \citenamefont {Marinelli}, \citenamefont {Moeller},\ and\ \citenamefont
    {Zemella}}]{Lutman.2016}%
    \BibitemOpen
    \bibfield  {author} {\bibinfo {author} {\bibfnamefont {A.~A.}\ \bibnamefont
    {Lutman}}, \bibinfo {author} {\bibfnamefont {T.~J.}\ \bibnamefont {Maxwell}},
    \bibinfo {author} {\bibfnamefont {J.~P.}\ \bibnamefont {MacArthur}}, \bibinfo
    {author} {\bibfnamefont {M.~W.}\ \bibnamefont {Guetg}}, \bibinfo {author}
    {\bibfnamefont {N.}~\bibnamefont {Berrah}}, \bibinfo {author} {\bibfnamefont
    {R.~N.}\ \bibnamefont {Coffee}}, \bibinfo {author} {\bibfnamefont
    {Y.}~\bibnamefont {Ding}}, \bibinfo {author} {\bibfnamefont {Z.}~\bibnamefont
    {Huang}}, \bibinfo {author} {\bibfnamefont {A.}~\bibnamefont {Marinelli}},
    \bibinfo {author} {\bibfnamefont {S.}~\bibnamefont {Moeller}}, \ and\
    \bibinfo {author} {\bibfnamefont {J.~C.}\ \bibnamefont {Zemella}},\ }\href
    {\doibase 10.1038/nphoton.2016.201} {\bibfield  {journal} {\bibinfo
    {journal} {Nat. Photonics}\ }\textbf {\bibinfo {volume} {10}},\ \bibinfo
    {pages} {745} (\bibinfo {year} {2016}{\natexlab{a}})}\BibitemShut {NoStop}%
  \bibitem [{\citenamefont {Zhang}\ \emph {et~al.}(2019)\citenamefont {Zhang},
    \citenamefont {Duris}, \citenamefont {Macarthur}, \citenamefont {Huang},\
    and\ \citenamefont {Marinelli}}]{Zhang.2019}%
    \BibitemOpen
    \bibfield  {author} {\bibinfo {author} {\bibfnamefont {Z.}~\bibnamefont
    {Zhang}}, \bibinfo {author} {\bibfnamefont {J.}~\bibnamefont {Duris}},
    \bibinfo {author} {\bibfnamefont {J.~P.}\ \bibnamefont {Macarthur}}, \bibinfo
    {author} {\bibfnamefont {Z.}~\bibnamefont {Huang}}, \ and\ \bibinfo {author}
    {\bibfnamefont {A.}~\bibnamefont {Marinelli}},\ }\href {\doibase
    10.1103/PhysRevAccelBeams.22.050701} {\bibfield  {journal} {\bibinfo
    {journal} {Phys. Rev. Accel. Beams}\ }\textbf {\bibinfo {volume} {22}},\
    \bibinfo {pages} {50701} (\bibinfo {year} {2019})}\BibitemShut {NoStop}%
  \bibitem [{\citenamefont {Yan}\ and\ \citenamefont {Deng}(2019)}]{Yan2019}%
    \BibitemOpen
    \bibfield  {author} {\bibinfo {author} {\bibfnamefont {J.}~\bibnamefont
    {Yan}}\ and\ \bibinfo {author} {\bibfnamefont {H.}~\bibnamefont {Deng}},\
    }\href {\doibase 10.1103/PhysRevAccelBeams.22.020703} {\bibfield  {journal}
    {\bibinfo  {journal} {Phys. Rev. Accel. Beams}\ }\textbf {\bibinfo {volume}
    {22}},\ \bibinfo {pages} {20703} (\bibinfo {year} {2019})}\BibitemShut
    {NoStop}%
  \bibitem [{\citenamefont {Deng}\ \emph {et~al.}(2019)\citenamefont {Deng},
    \citenamefont {Yan}, \citenamefont {Zhang}, \citenamefont {Sang},\ and\
    \citenamefont {Deng}}]{Deng:gb5079}%
    \BibitemOpen
    \bibfield  {author} {\bibinfo {author} {\bibfnamefont {B.}~\bibnamefont
    {Deng}}, \bibinfo {author} {\bibfnamefont {J.}~\bibnamefont {Yan}}, \bibinfo
    {author} {\bibfnamefont {Q.}~\bibnamefont {Zhang}}, \bibinfo {author}
    {\bibfnamefont {Y.}~\bibnamefont {Sang}}, \ and\ \bibinfo {author}
    {\bibfnamefont {H.}~\bibnamefont {Deng}},\ }\href {\doibase
    10.1107/S1600577518015175} {\bibfield  {journal} {\bibinfo  {journal} {J.
    Synchrotron Radiat.}\ }\textbf {\bibinfo {volume} {26}},\ \bibinfo {pages}
    {1} (\bibinfo {year} {2019})}\BibitemShut {NoStop}%
  \bibitem [{\citenamefont {Ferrari}\ \emph {et~al.}(2019)\citenamefont
    {Ferrari}, \citenamefont {Roussel}, \citenamefont {Buck}, \citenamefont
    {Callegari}, \citenamefont {Cucini}, \citenamefont {{De Ninno}},
    \citenamefont {Diviacco}, \citenamefont {Gauthier}, \citenamefont
    {Giannessi}, \citenamefont {Glaser}, \citenamefont {Hartmann}, \citenamefont
    {Penco}, \citenamefont {Scholz}, \citenamefont {Seltmann}, \citenamefont
    {Shevchuk}, \citenamefont {Viefhaus}, \citenamefont {Zangrando},\ and\
    \citenamefont {Allaria}}]{Ferrari.2019}%
    \BibitemOpen
    \bibfield  {author} {\bibinfo {author} {\bibfnamefont {E.}~\bibnamefont
    {Ferrari}}, \bibinfo {author} {\bibfnamefont {E.}~\bibnamefont {Roussel}},
    \bibinfo {author} {\bibfnamefont {J.}~\bibnamefont {Buck}}, \bibinfo {author}
    {\bibfnamefont {C.}~\bibnamefont {Callegari}}, \bibinfo {author}
    {\bibfnamefont {R.}~\bibnamefont {Cucini}}, \bibinfo {author} {\bibfnamefont
    {G.}~\bibnamefont {{De Ninno}}}, \bibinfo {author} {\bibfnamefont
    {B.}~\bibnamefont {Diviacco}}, \bibinfo {author} {\bibfnamefont
    {D.}~\bibnamefont {Gauthier}}, \bibinfo {author} {\bibfnamefont
    {L.}~\bibnamefont {Giannessi}}, \bibinfo {author} {\bibfnamefont
    {L.}~\bibnamefont {Glaser}}, \bibinfo {author} {\bibfnamefont
    {G.}~\bibnamefont {Hartmann}}, \bibinfo {author} {\bibfnamefont
    {G.}~\bibnamefont {Penco}}, \bibinfo {author} {\bibfnamefont
    {F.}~\bibnamefont {Scholz}}, \bibinfo {author} {\bibfnamefont
    {J.}~\bibnamefont {Seltmann}}, \bibinfo {author} {\bibfnamefont
    {I.}~\bibnamefont {Shevchuk}}, \bibinfo {author} {\bibfnamefont
    {J.}~\bibnamefont {Viefhaus}}, \bibinfo {author} {\bibfnamefont
    {M.}~\bibnamefont {Zangrando}}, \ and\ \bibinfo {author} {\bibfnamefont
    {E.~M.}\ \bibnamefont {Allaria}},\ }\href {\doibase
    10.1103/PhysRevAccelBeams.22.080701} {\bibfield  {journal} {\bibinfo
    {journal} {Phys. Rev. Accel. Beams}\ }\textbf {\bibinfo {volume} {22}},\
    \bibinfo {pages} {80701} (\bibinfo {year} {2019})}\BibitemShut {NoStop}%
  \bibitem [{\citenamefont {Lutman}\ \emph
    {et~al.}(2016{\natexlab{b}})\citenamefont {Lutman}, \citenamefont
    {MacArthur}, \citenamefont {Ilchen}, \citenamefont {Lindahl}, \citenamefont
    {Buck}, \citenamefont {Coffee}, \citenamefont {Dakovski}, \citenamefont
    {Dammann}, \citenamefont {Ding}, \citenamefont {D{\"{u}}rr}, \citenamefont
    {Glaser}, \citenamefont {Gr{\"{u}}nert}, \citenamefont {Hartmann},
    \citenamefont {Hartmann}, \citenamefont {Higley}, \citenamefont {Hirsch},
    \citenamefont {Levashov}, \citenamefont {Marinelli}, \citenamefont {Maxwell},
    \citenamefont {Mitra}, \citenamefont {Moeller}, \citenamefont {Osipov},
    \citenamefont {Peters}, \citenamefont {Planas}, \citenamefont {Shevchuk},
    \citenamefont {Schlotter}, \citenamefont {Scholz}, \citenamefont {Seltmann},
    \citenamefont {Viefhaus}, \citenamefont {Walter}, \citenamefont {Wolf},
    \citenamefont {Huang},\ and\ \citenamefont {Nuhn}}]{lutman2016polarization}%
    \BibitemOpen
    \bibfield  {author} {\bibinfo {author} {\bibfnamefont {A.~A.}\ \bibnamefont
    {Lutman}}, \bibinfo {author} {\bibfnamefont {J.~P.}\ \bibnamefont
    {MacArthur}}, \bibinfo {author} {\bibfnamefont {M.}~\bibnamefont {Ilchen}},
    \bibinfo {author} {\bibfnamefont {A.~O.}\ \bibnamefont {Lindahl}}, \bibinfo
    {author} {\bibfnamefont {J.}~\bibnamefont {Buck}}, \bibinfo {author}
    {\bibfnamefont {R.~N.}\ \bibnamefont {Coffee}}, \bibinfo {author}
    {\bibfnamefont {G.~L.}\ \bibnamefont {Dakovski}}, \bibinfo {author}
    {\bibfnamefont {L.}~\bibnamefont {Dammann}}, \bibinfo {author} {\bibfnamefont
    {Y.}~\bibnamefont {Ding}}, \bibinfo {author} {\bibfnamefont {H.~A.}\
    \bibnamefont {D{\"{u}}rr}}, \bibinfo {author} {\bibfnamefont
    {L.}~\bibnamefont {Glaser}}, \bibinfo {author} {\bibfnamefont
    {J.}~\bibnamefont {Gr{\"{u}}nert}}, \bibinfo {author} {\bibfnamefont
    {G.}~\bibnamefont {Hartmann}}, \bibinfo {author} {\bibfnamefont
    {N.}~\bibnamefont {Hartmann}}, \bibinfo {author} {\bibfnamefont
    {D.}~\bibnamefont {Higley}}, \bibinfo {author} {\bibfnamefont
    {K.}~\bibnamefont {Hirsch}}, \bibinfo {author} {\bibfnamefont {Y.~I.}\
    \bibnamefont {Levashov}}, \bibinfo {author} {\bibfnamefont {A.}~\bibnamefont
    {Marinelli}}, \bibinfo {author} {\bibfnamefont {T.}~\bibnamefont {Maxwell}},
    \bibinfo {author} {\bibfnamefont {A.}~\bibnamefont {Mitra}}, \bibinfo
    {author} {\bibfnamefont {S.}~\bibnamefont {Moeller}}, \bibinfo {author}
    {\bibfnamefont {T.}~\bibnamefont {Osipov}}, \bibinfo {author} {\bibfnamefont
    {F.}~\bibnamefont {Peters}}, \bibinfo {author} {\bibfnamefont
    {M.}~\bibnamefont {Planas}}, \bibinfo {author} {\bibfnamefont
    {I.}~\bibnamefont {Shevchuk}}, \bibinfo {author} {\bibfnamefont {W.~F.}\
    \bibnamefont {Schlotter}}, \bibinfo {author} {\bibfnamefont {F.}~\bibnamefont
    {Scholz}}, \bibinfo {author} {\bibfnamefont {J.}~\bibnamefont {Seltmann}},
    \bibinfo {author} {\bibfnamefont {J.}~\bibnamefont {Viefhaus}}, \bibinfo
    {author} {\bibfnamefont {P.}~\bibnamefont {Walter}}, \bibinfo {author}
    {\bibfnamefont {Z.~R.}\ \bibnamefont {Wolf}}, \bibinfo {author}
    {\bibfnamefont {Z.}~\bibnamefont {Huang}}, \ and\ \bibinfo {author}
    {\bibfnamefont {H.~D.}\ \bibnamefont {Nuhn}},\ }\href {\doibase
    10.1038/nphoton.2016.79} {\bibfield  {journal} {\bibinfo  {journal} {Nat.
    Photonics}\ }\textbf {\bibinfo {volume} {10}},\ \bibinfo {pages} {468}
    (\bibinfo {year} {2016}{\natexlab{b}})}\BibitemShut {NoStop}%
  \bibitem [{\citenamefont {Eisebitt}\ \emph {et~al.}(2004)\citenamefont
    {Eisebitt}, \citenamefont {L{\"{u}}ning}, \citenamefont {Schlotter},
    \citenamefont {L{\"{o}}rgen}, \citenamefont {Hellwig}, \citenamefont
    {Eberhardt},\ and\ \citenamefont {St{\"{o}}hr}}]{Eisebitt.2004}%
    \BibitemOpen
    \bibfield  {author} {\bibinfo {author} {\bibfnamefont {S.}~\bibnamefont
    {Eisebitt}}, \bibinfo {author} {\bibfnamefont {J.}~\bibnamefont
    {L{\"{u}}ning}}, \bibinfo {author} {\bibfnamefont {W.~F.}\ \bibnamefont
    {Schlotter}}, \bibinfo {author} {\bibfnamefont {M.}~\bibnamefont
    {L{\"{o}}rgen}}, \bibinfo {author} {\bibfnamefont {O.}~\bibnamefont
    {Hellwig}}, \bibinfo {author} {\bibfnamefont {W.}~\bibnamefont {Eberhardt}},
    \ and\ \bibinfo {author} {\bibfnamefont {J.}~\bibnamefont {St{\"{o}}hr}},\
    }\href {\doibase 10.1038/nature03139} {\bibfield  {journal} {\bibinfo
    {journal} {Nature}\ }\textbf {\bibinfo {volume} {432}},\ \bibinfo {pages}
    {885} (\bibinfo {year} {2004})}\BibitemShut {NoStop}%
  \bibitem [{\citenamefont {Boeglin}\ \emph {et~al.}(2010)\citenamefont
    {Boeglin}, \citenamefont {Beaurepaire}, \citenamefont {Halt{\'{e}}},
    \citenamefont {L{\'{o}}pez-Flores}, \citenamefont {Stamm}, \citenamefont
    {Pontius}, \citenamefont {D{\"{u}}rr},\ and\ \citenamefont
    {Bigot}}]{Boeglin.2010}%
    \BibitemOpen
    \bibfield  {author} {\bibinfo {author} {\bibfnamefont {C.}~\bibnamefont
    {Boeglin}}, \bibinfo {author} {\bibfnamefont {E.}~\bibnamefont
    {Beaurepaire}}, \bibinfo {author} {\bibfnamefont {V.}~\bibnamefont
    {Halt{\'{e}}}}, \bibinfo {author} {\bibfnamefont {V.}~\bibnamefont
    {L{\'{o}}pez-Flores}}, \bibinfo {author} {\bibfnamefont {C.}~\bibnamefont
    {Stamm}}, \bibinfo {author} {\bibfnamefont {N.}~\bibnamefont {Pontius}},
    \bibinfo {author} {\bibfnamefont {H.~A.}\ \bibnamefont {D{\"{u}}rr}}, \ and\
    \bibinfo {author} {\bibfnamefont {J.~Y.}\ \bibnamefont {Bigot}},\ }\href
    {\doibase 10.1038/nature09070} {\bibfield  {journal} {\bibinfo  {journal}
    {Nature}\ }\textbf {\bibinfo {volume} {465}},\ \bibinfo {pages} {458}
    (\bibinfo {year} {2010})}\BibitemShut {NoStop}%
  \bibitem [{\citenamefont {Inami}(2017)}]{Toshiya.2017}%
    \BibitemOpen
    \bibfield  {author} {\bibinfo {author} {\bibfnamefont {T.}~\bibnamefont
    {Inami}},\ }\href {\doibase 10.1103/PhysRevLett.119.137203} {\bibfield
    {journal} {\bibinfo  {journal} {Phys. Rev. Lett.}\ }\textbf {\bibinfo
    {volume} {119}},\ \bibinfo {pages} {137203} (\bibinfo {year}
    {2017})}\BibitemShut {NoStop}%
  \bibitem [{\citenamefont {Sato}\ \emph {et~al.}(2002)\citenamefont {Sato},
    \citenamefont {Ueji}, \citenamefont {Okitsu}, \citenamefont {Matsushita},
    \citenamefont {Saito}, \citenamefont {Takayama},\ and\ \citenamefont
    {Amemiya}}]{K_Sato.2010}%
    \BibitemOpen
    \bibfield  {author} {\bibinfo {author} {\bibfnamefont {K.}~\bibnamefont
    {Sato}}, \bibinfo {author} {\bibfnamefont {Y.}~\bibnamefont {Ueji}}, \bibinfo
    {author} {\bibfnamefont {K.}~\bibnamefont {Okitsu}}, \bibinfo {author}
    {\bibfnamefont {T.}~\bibnamefont {Matsushita}}, \bibinfo {author}
    {\bibfnamefont {J.}~\bibnamefont {Saito}}, \bibinfo {author} {\bibfnamefont
    {T.}~\bibnamefont {Takayama}}, \ and\ \bibinfo {author} {\bibfnamefont
    {Y.}~\bibnamefont {Amemiya}},\ }\href {\doibase 10.3379/tmjpn2001.2.238}
    {\bibfield  {journal} {\bibinfo  {journal} {Trans. Magn. Soc. Japan}\
    }\textbf {\bibinfo {volume} {2}},\ \bibinfo {pages} {238} (\bibinfo {year}
    {2002})}\BibitemShut {NoStop}%
  \bibitem [{\citenamefont {Fujiwara}\ \emph {et~al.}(2016)\citenamefont
    {Fujiwara}, \citenamefont {Naimen}, \citenamefont {Higashiya}, \citenamefont
    {Kanai}, \citenamefont {Yomosa}, \citenamefont {Yamagami}, \citenamefont
    {Kiss}, \citenamefont {Kadono}, \citenamefont {Imada}, \citenamefont
    {Yamasaki}, \citenamefont {Takase}, \citenamefont {Otsuka}, \citenamefont
    {Shimizu}, \citenamefont {Shingubara}, \citenamefont {Suga}, \citenamefont
    {Yabashi}, \citenamefont {Tamasaku}, \citenamefont {Ishikawa},\ and\
    \citenamefont {Sekiyama}}]{Fujiwara.2016}%
    \BibitemOpen
    \bibfield  {author} {\bibinfo {author} {\bibfnamefont {H.}~\bibnamefont
    {Fujiwara}}, \bibinfo {author} {\bibfnamefont {S.}~\bibnamefont {Naimen}},
    \bibinfo {author} {\bibfnamefont {A.}~\bibnamefont {Higashiya}}, \bibinfo
    {author} {\bibfnamefont {Y.}~\bibnamefont {Kanai}}, \bibinfo {author}
    {\bibfnamefont {H.}~\bibnamefont {Yomosa}}, \bibinfo {author} {\bibfnamefont
    {K.}~\bibnamefont {Yamagami}}, \bibinfo {author} {\bibfnamefont
    {T.}~\bibnamefont {Kiss}}, \bibinfo {author} {\bibfnamefont {T.}~\bibnamefont
    {Kadono}}, \bibinfo {author} {\bibfnamefont {S.}~\bibnamefont {Imada}},
    \bibinfo {author} {\bibfnamefont {A.}~\bibnamefont {Yamasaki}}, \bibinfo
    {author} {\bibfnamefont {K.}~\bibnamefont {Takase}}, \bibinfo {author}
    {\bibfnamefont {S.}~\bibnamefont {Otsuka}}, \bibinfo {author} {\bibfnamefont
    {T.}~\bibnamefont {Shimizu}}, \bibinfo {author} {\bibfnamefont
    {S.}~\bibnamefont {Shingubara}}, \bibinfo {author} {\bibfnamefont
    {S.}~\bibnamefont {Suga}}, \bibinfo {author} {\bibfnamefont {M.}~\bibnamefont
    {Yabashi}}, \bibinfo {author} {\bibfnamefont {K.}~\bibnamefont {Tamasaku}},
    \bibinfo {author} {\bibfnamefont {T.}~\bibnamefont {Ishikawa}}, \ and\
    \bibinfo {author} {\bibfnamefont {A.}~\bibnamefont {Sekiyama}},\ }\href
    {\doibase 10.1107/S1600577516003003} {\bibfield  {journal} {\bibinfo
    {journal} {J. Synchrotron Radiat.}\ }\textbf {\bibinfo {volume} {23}},\
    \bibinfo {pages} {735} (\bibinfo {year} {2016})},\ \Eprint
    {http://arxiv.org/abs/1505.04591} {arXiv:1505.04591} \BibitemShut {NoStop}%
  \bibitem [{\citenamefont {Logan}\ \emph {et~al.}(2016)\citenamefont {Logan},
    \citenamefont {Harder}, \citenamefont {Li}, \citenamefont {Haskel},
    \citenamefont {Chen}, \citenamefont {Winarski}, \citenamefont {Fuesz},
    \citenamefont {Schlagel}, \citenamefont {Vine}, \citenamefont {Benson},\ and\
    \citenamefont {McNulty}}]{Logan.2016}%
    \BibitemOpen
    \bibfield  {author} {\bibinfo {author} {\bibfnamefont {J.}~\bibnamefont
    {Logan}}, \bibinfo {author} {\bibfnamefont {R.}~\bibnamefont {Harder}},
    \bibinfo {author} {\bibfnamefont {L.}~\bibnamefont {Li}}, \bibinfo {author}
    {\bibfnamefont {D.}~\bibnamefont {Haskel}}, \bibinfo {author} {\bibfnamefont
    {P.}~\bibnamefont {Chen}}, \bibinfo {author} {\bibfnamefont {R.}~\bibnamefont
    {Winarski}}, \bibinfo {author} {\bibfnamefont {P.}~\bibnamefont {Fuesz}},
    \bibinfo {author} {\bibfnamefont {D.}~\bibnamefont {Schlagel}}, \bibinfo
    {author} {\bibfnamefont {D.}~\bibnamefont {Vine}}, \bibinfo {author}
    {\bibfnamefont {C.}~\bibnamefont {Benson}}, \ and\ \bibinfo {author}
    {\bibfnamefont {I.}~\bibnamefont {McNulty}},\ }\href {\doibase
    10.1107/S1600577516009632} {\bibfield  {journal} {\bibinfo  {journal} {J.
    Synchrotron Radiat.}\ }\textbf {\bibinfo {volume} {23}},\ \bibinfo {pages}
    {1210} (\bibinfo {year} {2016})}\BibitemShut {NoStop}%
  \bibitem [{\citenamefont {Suzuki}(2014)}]{Suzuki.2014}%
    \BibitemOpen
    \bibfield  {author} {\bibinfo {author} {\bibfnamefont {M.}~\bibnamefont
    {Suzuki}},\ }\href {\doibase 10.1093/jmicro/dfu041} {\bibfield  {journal}
    {\bibinfo  {journal} {Microscopy}\ }\textbf {\bibinfo {volume} {63}},\
    \bibinfo {pages} {i14.1} (\bibinfo {year} {2014})}\BibitemShut {NoStop}%
  \bibitem [{\citenamefont {Shen}\ \emph {et~al.}(2018)\citenamefont {Shen},
    \citenamefont {Bu}, \citenamefont {Xu}, \citenamefont {Xu}, \citenamefont
    {Ji}, \citenamefont {Li},\ and\ \citenamefont {Xu}}]{Shen.2018}%
    \BibitemOpen
    \bibfield  {author} {\bibinfo {author} {\bibfnamefont {B.}~\bibnamefont
    {Shen}}, \bibinfo {author} {\bibfnamefont {Z.}~\bibnamefont {Bu}}, \bibinfo
    {author} {\bibfnamefont {J.}~\bibnamefont {Xu}}, \bibinfo {author}
    {\bibfnamefont {T.}~\bibnamefont {Xu}}, \bibinfo {author} {\bibfnamefont
    {L.}~\bibnamefont {Ji}}, \bibinfo {author} {\bibfnamefont {R.}~\bibnamefont
    {Li}}, \ and\ \bibinfo {author} {\bibfnamefont {Z.}~\bibnamefont {Xu}},\
    }\href {\doibase 10.1088/1361-6587/aaa7fb} {\bibfield  {journal} {\bibinfo
    {journal} {Plasma Phys. Control. Fusion}\ }\textbf {\bibinfo {volume} {60}},\
    \bibinfo {pages} {44002} (\bibinfo {year} {2018})}\BibitemShut {NoStop}%
  \bibitem [{\citenamefont {Temnykh}(2008)}]{Temnykh.2008}%
    \BibitemOpen
    \bibfield  {author} {\bibinfo {author} {\bibfnamefont {A.~B.}\ \bibnamefont
    {Temnykh}},\ }\href {\doibase 10.1103/PhysRevSTAB.11.120702} {\bibfield
    {journal} {\bibinfo  {journal} {Phys. Rev. Spec. Top. - Accel. Beams}\
    }\textbf {\bibinfo {volume} {11}},\ \bibinfo {pages} {235} (\bibinfo {year}
    {2008})}\BibitemShut {NoStop}%
  \bibitem [{\citenamefont {Sasaki}(1994)}]{Sasaki.1994}%
    \BibitemOpen
    \bibfield  {author} {\bibinfo {author} {\bibfnamefont {S.}~\bibnamefont
    {Sasaki}},\ }\href {\doibase 10.1016/0168-9002(94)91859-7} {\bibfield
    {journal} {\bibinfo  {journal} {Nucl. Inst. Methods Phys. Res. A}\ }\textbf
    {\bibinfo {volume} {347}},\ \bibinfo {pages} {83} (\bibinfo {year}
    {1994})}\BibitemShut {NoStop}%
  \bibitem [{\citenamefont {Spezzani}\ \emph {et~al.}(2011)\citenamefont
    {Spezzani}, \citenamefont {Allaria}, \citenamefont {Coreno}, \citenamefont
    {Diviacco}, \citenamefont {Ferrari}, \citenamefont {Geloni}, \citenamefont
    {Karantzoulis}, \citenamefont {Mahieu}, \citenamefont {Vento},\ and\
    \citenamefont {{De Ninno}}}]{Spezzani.2011}%
    \BibitemOpen
    \bibfield  {author} {\bibinfo {author} {\bibfnamefont {C.}~\bibnamefont
    {Spezzani}}, \bibinfo {author} {\bibfnamefont {E.}~\bibnamefont {Allaria}},
    \bibinfo {author} {\bibfnamefont {M.}~\bibnamefont {Coreno}}, \bibinfo
    {author} {\bibfnamefont {B.}~\bibnamefont {Diviacco}}, \bibinfo {author}
    {\bibfnamefont {E.}~\bibnamefont {Ferrari}}, \bibinfo {author} {\bibfnamefont
    {G.}~\bibnamefont {Geloni}}, \bibinfo {author} {\bibfnamefont
    {E.}~\bibnamefont {Karantzoulis}}, \bibinfo {author} {\bibfnamefont
    {B.}~\bibnamefont {Mahieu}}, \bibinfo {author} {\bibfnamefont
    {M.}~\bibnamefont {Vento}}, \ and\ \bibinfo {author} {\bibfnamefont
    {G.}~\bibnamefont {{De Ninno}}},\ }\href {\doibase
    10.1103/PhysRevLett.107.084801} {\bibfield  {journal} {\bibinfo  {journal}
    {Phys. Rev. Lett.}\ }\textbf {\bibinfo {volume} {107}},\ \bibinfo {pages}
    {84801} (\bibinfo {year} {2011})}\BibitemShut {NoStop}%
  \bibitem [{\citenamefont {Allaria}\ \emph {et~al.}(2014)\citenamefont
    {Allaria}, \citenamefont {Diviacco}, \citenamefont {Callegari}, \citenamefont
    {Finetti}, \citenamefont {Mahieu}, \citenamefont {Viefhaus}, \citenamefont
    {Zangrando}, \citenamefont {{De Ninno}}, \citenamefont {Lambert},
    \citenamefont {Ferrari}, \citenamefont {Buck}, \citenamefont {Ilchen},
    \citenamefont {Vodungbo}, \citenamefont {Mahne}, \citenamefont {Svetina},
    \citenamefont {Spezzani}, \citenamefont {{Di Mitri}}, \citenamefont {Penco},
    \citenamefont {Trov{\'{o}}}, \citenamefont {Fawley}, \citenamefont
    {Rebernik}, \citenamefont {Gauthier}, \citenamefont {Grazioli}, \citenamefont
    {Coreno}, \citenamefont {Ressel}, \citenamefont {Kivim{\"{a}}ki},
    \citenamefont {Mazza}, \citenamefont {Glaser}, \citenamefont {Scholz},
    \citenamefont {Seltmann}, \citenamefont {Gessler}, \citenamefont
    {Gr{\"{u}}nert}, \citenamefont {{De Fanis}}, \citenamefont {Meyer},
    \citenamefont {Knie}, \citenamefont {Moeller}, \citenamefont {Raimondi},
    \citenamefont {Capotondi}, \citenamefont {Pedersoli}, \citenamefont {Plekan},
    \citenamefont {Danailov}, \citenamefont {Demidovich}, \citenamefont
    {Nikolov}, \citenamefont {Abrami}, \citenamefont {Gautier}, \citenamefont
    {L{\"{u}}ning}, \citenamefont {Zeitoun},\ and\ \citenamefont
    {Giannessi}}]{Allaria.2014}%
    \BibitemOpen
    \bibfield  {author} {\bibinfo {author} {\bibfnamefont {E.}~\bibnamefont
    {Allaria}}, \bibinfo {author} {\bibfnamefont {B.}~\bibnamefont {Diviacco}},
    \bibinfo {author} {\bibfnamefont {C.}~\bibnamefont {Callegari}}, \bibinfo
    {author} {\bibfnamefont {P.}~\bibnamefont {Finetti}}, \bibinfo {author}
    {\bibfnamefont {B.}~\bibnamefont {Mahieu}}, \bibinfo {author} {\bibfnamefont
    {J.}~\bibnamefont {Viefhaus}}, \bibinfo {author} {\bibfnamefont
    {M.}~\bibnamefont {Zangrando}}, \bibinfo {author} {\bibfnamefont
    {G.}~\bibnamefont {{De Ninno}}}, \bibinfo {author} {\bibfnamefont
    {G.}~\bibnamefont {Lambert}}, \bibinfo {author} {\bibfnamefont
    {E.}~\bibnamefont {Ferrari}}, \bibinfo {author} {\bibfnamefont
    {J.}~\bibnamefont {Buck}}, \bibinfo {author} {\bibfnamefont {M.}~\bibnamefont
    {Ilchen}}, \bibinfo {author} {\bibfnamefont {B.}~\bibnamefont {Vodungbo}},
    \bibinfo {author} {\bibfnamefont {N.}~\bibnamefont {Mahne}}, \bibinfo
    {author} {\bibfnamefont {C.}~\bibnamefont {Svetina}}, \bibinfo {author}
    {\bibfnamefont {C.}~\bibnamefont {Spezzani}}, \bibinfo {author}
    {\bibfnamefont {S.}~\bibnamefont {{Di Mitri}}}, \bibinfo {author}
    {\bibfnamefont {G.}~\bibnamefont {Penco}}, \bibinfo {author} {\bibfnamefont
    {M.}~\bibnamefont {Trov{\'{o}}}}, \bibinfo {author} {\bibfnamefont {W.~M.}\
    \bibnamefont {Fawley}}, \bibinfo {author} {\bibfnamefont {P.~R.}\
    \bibnamefont {Rebernik}}, \bibinfo {author} {\bibfnamefont {D.}~\bibnamefont
    {Gauthier}}, \bibinfo {author} {\bibfnamefont {C.}~\bibnamefont {Grazioli}},
    \bibinfo {author} {\bibfnamefont {M.}~\bibnamefont {Coreno}}, \bibinfo
    {author} {\bibfnamefont {B.}~\bibnamefont {Ressel}}, \bibinfo {author}
    {\bibfnamefont {A.}~\bibnamefont {Kivim{\"{a}}ki}}, \bibinfo {author}
    {\bibfnamefont {T.}~\bibnamefont {Mazza}}, \bibinfo {author} {\bibfnamefont
    {L.}~\bibnamefont {Glaser}}, \bibinfo {author} {\bibfnamefont
    {F.}~\bibnamefont {Scholz}}, \bibinfo {author} {\bibfnamefont
    {J.}~\bibnamefont {Seltmann}}, \bibinfo {author} {\bibfnamefont
    {P.}~\bibnamefont {Gessler}}, \bibinfo {author} {\bibfnamefont
    {J.}~\bibnamefont {Gr{\"{u}}nert}}, \bibinfo {author} {\bibfnamefont
    {A.}~\bibnamefont {{De Fanis}}}, \bibinfo {author} {\bibfnamefont
    {M.}~\bibnamefont {Meyer}}, \bibinfo {author} {\bibfnamefont
    {A.}~\bibnamefont {Knie}}, \bibinfo {author} {\bibfnamefont {S.~P.}\
    \bibnamefont {Moeller}}, \bibinfo {author} {\bibfnamefont {L.}~\bibnamefont
    {Raimondi}}, \bibinfo {author} {\bibfnamefont {F.}~\bibnamefont {Capotondi}},
    \bibinfo {author} {\bibfnamefont {E.}~\bibnamefont {Pedersoli}}, \bibinfo
    {author} {\bibfnamefont {O.}~\bibnamefont {Plekan}}, \bibinfo {author}
    {\bibfnamefont {M.~B.}\ \bibnamefont {Danailov}}, \bibinfo {author}
    {\bibfnamefont {A.}~\bibnamefont {Demidovich}}, \bibinfo {author}
    {\bibfnamefont {I.}~\bibnamefont {Nikolov}}, \bibinfo {author} {\bibfnamefont
    {A.}~\bibnamefont {Abrami}}, \bibinfo {author} {\bibfnamefont
    {J.}~\bibnamefont {Gautier}}, \bibinfo {author} {\bibfnamefont
    {J.}~\bibnamefont {L{\"{u}}ning}}, \bibinfo {author} {\bibfnamefont
    {P.}~\bibnamefont {Zeitoun}}, \ and\ \bibinfo {author} {\bibfnamefont
    {L.}~\bibnamefont {Giannessi}},\ }\href {\doibase 10.1103/PhysRevX.4.041040}
    {\bibfield  {journal} {\bibinfo  {journal} {Phys. Rev. X}\ }\textbf {\bibinfo
    {volume} {4}},\ \bibinfo {pages} {1} (\bibinfo {year} {2014})}\BibitemShut
    {NoStop}%
  \bibitem [{\citenamefont {Kim}(1984)}]{Kim.1984}%
    \BibitemOpen
    \bibfield  {author} {\bibinfo {author} {\bibfnamefont {K.~J.}\ \bibnamefont
    {Kim}},\ }\href {\doibase 10.1016/0167-5087(84)90354-5} {\bibfield  {journal}
    {\bibinfo  {journal} {Nucl. Instruments Methods Phys. Res.}\ }\textbf
    {\bibinfo {volume} {219}},\ \bibinfo {pages} {425} (\bibinfo {year}
    {1984})}\BibitemShut {NoStop}%
  \bibitem [{\citenamefont {Kim}(2000)}]{Kim.2000}%
    \BibitemOpen
    \bibfield  {author} {\bibinfo {author} {\bibfnamefont {K.~J.}\ \bibnamefont
    {Kim}},\ }\href {\doibase 10.1016/S0168-9002(00)00137-6} {\bibfield
    {journal} {\bibinfo  {journal} {Nucl. Instruments Methods Phys. Res. Sect. A
    Accel. Spectrometers, Detect. Assoc. Equip.}\ }\textbf {\bibinfo {volume}
    {445}},\ \bibinfo {pages} {329} (\bibinfo {year} {2000})}\BibitemShut
    {NoStop}%
  \bibitem [{\citenamefont {Deng}\ \emph {et~al.}(2014)\citenamefont {Deng},
    \citenamefont {Zhang}, \citenamefont {Feng}, \citenamefont {Feng},
    \citenamefont {Liu}, \citenamefont {Wang}, \citenamefont {Lan}, \citenamefont
    {Wang}, \citenamefont {Zhang}, \citenamefont {Liu}, \citenamefont {Chen},
    \citenamefont {Zhang}, \citenamefont {Lin}, \citenamefont {Zhang},
    \citenamefont {Wang},\ and\ \citenamefont {Zhao}}]{Deng.2014}%
    \BibitemOpen
    \bibfield  {author} {\bibinfo {author} {\bibfnamefont {H.}~\bibnamefont
    {Deng}}, \bibinfo {author} {\bibfnamefont {T.}~\bibnamefont {Zhang}},
    \bibinfo {author} {\bibfnamefont {L.}~\bibnamefont {Feng}}, \bibinfo {author}
    {\bibfnamefont {C.}~\bibnamefont {Feng}}, \bibinfo {author} {\bibfnamefont
    {B.}~\bibnamefont {Liu}}, \bibinfo {author} {\bibfnamefont {X.}~\bibnamefont
    {Wang}}, \bibinfo {author} {\bibfnamefont {T.}~\bibnamefont {Lan}}, \bibinfo
    {author} {\bibfnamefont {G.}~\bibnamefont {Wang}}, \bibinfo {author}
    {\bibfnamefont {W.}~\bibnamefont {Zhang}}, \bibinfo {author} {\bibfnamefont
    {X.}~\bibnamefont {Liu}}, \bibinfo {author} {\bibfnamefont {J.}~\bibnamefont
    {Chen}}, \bibinfo {author} {\bibfnamefont {M.}~\bibnamefont {Zhang}},
    \bibinfo {author} {\bibfnamefont {G.}~\bibnamefont {Lin}}, \bibinfo {author}
    {\bibfnamefont {M.}~\bibnamefont {Zhang}}, \bibinfo {author} {\bibfnamefont
    {D.}~\bibnamefont {Wang}}, \ and\ \bibinfo {author} {\bibfnamefont
    {Z.}~\bibnamefont {Zhao}},\ }\href {\doibase 10.1103/PhysRevSTAB.17.020704}
    {\bibfield  {journal} {\bibinfo  {journal} {Phys. Rev. Spec. Top. - Accel.
    Beams}\ }\textbf {\bibinfo {volume} {17}} (\bibinfo {year} {2014}),\
    10.1103/PhysRevSTAB.17.020704}\BibitemShut {NoStop}%
  \bibitem [{\citenamefont {Ding}\ and\ \citenamefont {Huang}(2008)}]{Ding.2008}%
    \BibitemOpen
    \bibfield  {author} {\bibinfo {author} {\bibfnamefont {Y.}~\bibnamefont
    {Ding}}\ and\ \bibinfo {author} {\bibfnamefont {Z.}~\bibnamefont {Huang}},\
    }\href {\doibase 10.1103/PhysRevSTAB.11.030702} {\bibfield  {journal}
    {\bibinfo  {journal} {Phys. Rev. Spec. Top. - Accel. Beams}\ }\textbf
    {\bibinfo {volume} {11}} (\bibinfo {year} {2008}),\
    10.1103/PhysRevSTAB.11.030702}\BibitemShut {NoStop}%
  \bibitem [{\citenamefont {Kim}\ \emph {et~al.}(2008)\citenamefont {Kim},
    \citenamefont {Shvyd'ko},\ and\ \citenamefont {Reiche}}]{kim2008proposal}%
    \BibitemOpen
    \bibfield  {author} {\bibinfo {author} {\bibfnamefont {K.~J.}\ \bibnamefont
    {Kim}}, \bibinfo {author} {\bibfnamefont {Y.}~\bibnamefont {Shvyd'ko}}, \
    and\ \bibinfo {author} {\bibfnamefont {S.}~\bibnamefont {Reiche}},\ }\href
    {\doibase 10.1103/PhysRevLett.100.244802} {\bibfield  {journal} {\bibinfo
    {journal} {Phys. Rev. Lett.}\ }\textbf {\bibinfo {volume} {100}},\ \bibinfo
    {pages} {244802} (\bibinfo {year} {2008})}\BibitemShut {NoStop}%
  \bibitem [{\citenamefont {Li}\ and\ \citenamefont {Deng}(2018)}]{GainLikai}%
    \BibitemOpen
    \bibfield  {author} {\bibinfo {author} {\bibfnamefont {K.}~\bibnamefont
    {Li}}\ and\ \bibinfo {author} {\bibfnamefont {H.}~\bibnamefont {Deng}},\
    }\href {\doibase 10.1063/1.5037180} {\bibfield  {journal} {\bibinfo
    {journal} {Appl. Phys. Lett.}\ }\textbf {\bibinfo {volume} {113}},\ \bibinfo
    {pages} {61106} (\bibinfo {year} {2018})}\BibitemShut {NoStop}%
  \bibitem [{\citenamefont {Wang}\ \emph {et~al.}(2012)\citenamefont {Wang},
    \citenamefont {Zhu}, \citenamefont {Wu}, \citenamefont {Graves},
    \citenamefont {Schaffert}, \citenamefont {Rander}, \citenamefont
    {M{\"{u}}ller}, \citenamefont {Vodungbo}, \citenamefont {Baumier},
    \citenamefont {Bernstein}, \citenamefont {Br{\"{a}}uer}, \citenamefont
    {Cros}, \citenamefont {{De Jong}}, \citenamefont {Delaunay}, \citenamefont
    {Fognini}, \citenamefont {Kukreja}, \citenamefont {Lee}, \citenamefont
    {L{\'{o}}pez-Flores}, \citenamefont {Mohanty}, \citenamefont {Pfau},
    \citenamefont {Popescu}, \citenamefont {Sacchi}, \citenamefont {Sardinha},
    \citenamefont {Sirotti}, \citenamefont {Zeitoun}, \citenamefont
    {Messerschmidt}, \citenamefont {Turner}, \citenamefont {Schlotter},
    \citenamefont {Hellwig}, \citenamefont {Mattana}, \citenamefont {Jaouen},
    \citenamefont {Fortuna}, \citenamefont {Acremann}, \citenamefont {Gutt},
    \citenamefont {D{\"{u}}rr}, \citenamefont {Beaurepaire}, \citenamefont
    {Boeglin}, \citenamefont {Eisebitt}, \citenamefont {Gr{\"{u}}bel},
    \citenamefont {L{\"{u}}ning}, \citenamefont {St{\"{o}}hr},\ and\
    \citenamefont {Scherz}}]{Wang.2012}%
    \BibitemOpen
    \bibfield  {author} {\bibinfo {author} {\bibfnamefont {T.}~\bibnamefont
    {Wang}}, \bibinfo {author} {\bibfnamefont {D.}~\bibnamefont {Zhu}}, \bibinfo
    {author} {\bibfnamefont {B.}~\bibnamefont {Wu}}, \bibinfo {author}
    {\bibfnamefont {C.}~\bibnamefont {Graves}}, \bibinfo {author} {\bibfnamefont
    {S.}~\bibnamefont {Schaffert}}, \bibinfo {author} {\bibfnamefont
    {T.}~\bibnamefont {Rander}}, \bibinfo {author} {\bibfnamefont
    {L.}~\bibnamefont {M{\"{u}}ller}}, \bibinfo {author} {\bibfnamefont
    {B.}~\bibnamefont {Vodungbo}}, \bibinfo {author} {\bibfnamefont
    {C.}~\bibnamefont {Baumier}}, \bibinfo {author} {\bibfnamefont {D.~P.}\
    \bibnamefont {Bernstein}}, \bibinfo {author} {\bibfnamefont {B.}~\bibnamefont
    {Br{\"{a}}uer}}, \bibinfo {author} {\bibfnamefont {V.}~\bibnamefont {Cros}},
    \bibinfo {author} {\bibfnamefont {S.}~\bibnamefont {{De Jong}}}, \bibinfo
    {author} {\bibfnamefont {R.}~\bibnamefont {Delaunay}}, \bibinfo {author}
    {\bibfnamefont {A.}~\bibnamefont {Fognini}}, \bibinfo {author} {\bibfnamefont
    {R.}~\bibnamefont {Kukreja}}, \bibinfo {author} {\bibfnamefont
    {S.}~\bibnamefont {Lee}}, \bibinfo {author} {\bibfnamefont {V.}~\bibnamefont
    {L{\'{o}}pez-Flores}}, \bibinfo {author} {\bibfnamefont {J.}~\bibnamefont
    {Mohanty}}, \bibinfo {author} {\bibfnamefont {B.}~\bibnamefont {Pfau}},
    \bibinfo {author} {\bibfnamefont {H.}~\bibnamefont {Popescu}}, \bibinfo
    {author} {\bibfnamefont {M.}~\bibnamefont {Sacchi}}, \bibinfo {author}
    {\bibfnamefont {A.~B.}\ \bibnamefont {Sardinha}}, \bibinfo {author}
    {\bibfnamefont {F.}~\bibnamefont {Sirotti}}, \bibinfo {author} {\bibfnamefont
    {P.}~\bibnamefont {Zeitoun}}, \bibinfo {author} {\bibfnamefont
    {M.}~\bibnamefont {Messerschmidt}}, \bibinfo {author} {\bibfnamefont {J.~J.}\
    \bibnamefont {Turner}}, \bibinfo {author} {\bibfnamefont {W.~F.}\
    \bibnamefont {Schlotter}}, \bibinfo {author} {\bibfnamefont {O.}~\bibnamefont
    {Hellwig}}, \bibinfo {author} {\bibfnamefont {R.}~\bibnamefont {Mattana}},
    \bibinfo {author} {\bibfnamefont {N.}~\bibnamefont {Jaouen}}, \bibinfo
    {author} {\bibfnamefont {F.}~\bibnamefont {Fortuna}}, \bibinfo {author}
    {\bibfnamefont {Y.}~\bibnamefont {Acremann}}, \bibinfo {author}
    {\bibfnamefont {C.}~\bibnamefont {Gutt}}, \bibinfo {author} {\bibfnamefont
    {H.~A.}\ \bibnamefont {D{\"{u}}rr}}, \bibinfo {author} {\bibfnamefont
    {E.}~\bibnamefont {Beaurepaire}}, \bibinfo {author} {\bibfnamefont
    {C.}~\bibnamefont {Boeglin}}, \bibinfo {author} {\bibfnamefont
    {S.}~\bibnamefont {Eisebitt}}, \bibinfo {author} {\bibfnamefont
    {G.}~\bibnamefont {Gr{\"{u}}bel}}, \bibinfo {author} {\bibfnamefont
    {J.}~\bibnamefont {L{\"{u}}ning}}, \bibinfo {author} {\bibfnamefont
    {J.}~\bibnamefont {St{\"{o}}hr}}, \ and\ \bibinfo {author} {\bibfnamefont
    {A.~O.}\ \bibnamefont {Scherz}},\ }\href {\doibase
    10.1103/PhysRevLett.108.267403} {\bibfield  {journal} {\bibinfo  {journal}
    {Phys. Rev. Lett.}\ }\textbf {\bibinfo {volume} {108}},\ \bibinfo {pages}
    {267403} (\bibinfo {year} {2012})}\BibitemShut {NoStop}%
  \bibitem [{\citenamefont {Kim}\ \emph {et~al.}(2017)\citenamefont {Kim},
    \citenamefont {Huang},\ and\ \citenamefont {Lindberg}}]{Kim.2017}%
    \BibitemOpen
    \bibfield  {author} {\bibinfo {author} {\bibfnamefont {K.-J.}\ \bibnamefont
    {Kim}}, \bibinfo {author} {\bibfnamefont {Z.}~\bibnamefont {Huang}}, \ and\
    \bibinfo {author} {\bibfnamefont {R.}~\bibnamefont {Lindberg}},\ }\href@noop
    {} {\emph {\bibinfo {title} {{Synchrotron radiation and free-electron lasers:
    Principles of coherent x-ray generation / Kwang-Je Kim, Zhirong Huang, Ryan
    Lindberg}}}}\ (\bibinfo  {publisher} {Cambridge University Press},\ \bibinfo
    {address} {Cambridge},\ \bibinfo {year} {2017})\BibitemShut {NoStop}%
  \bibitem [{\citenamefont {Li}\ \emph {et~al.}(2017)\citenamefont {Li},
    \citenamefont {Song},\ and\ \citenamefont {Deng}}]{li2017simplified}%
    \BibitemOpen
    \bibfield  {author} {\bibinfo {author} {\bibfnamefont {K.}~\bibnamefont
    {Li}}, \bibinfo {author} {\bibfnamefont {M.}~\bibnamefont {Song}}, \ and\
    \bibinfo {author} {\bibfnamefont {H.}~\bibnamefont {Deng}},\ }\href {\doibase
    10.1103/PhysRevAccelBeams.20.030702} {\bibfield  {journal} {\bibinfo
    {journal} {Phys. Rev. Accel. Beams}\ }\textbf {\bibinfo {volume} {20}},\
    \bibinfo {pages} {30702} (\bibinfo {year} {2017})}\BibitemShut {NoStop}%
  \bibitem [{\citenamefont {Kroll}\ \emph {et~al.}(1981)\citenamefont {Kroll},
    \citenamefont {Morton},\ and\ \citenamefont {Rosenbluth}}]{Kroll.1981}%
    \BibitemOpen
    \bibfield  {author} {\bibinfo {author} {\bibfnamefont {N.~M.}\ \bibnamefont
    {Kroll}}, \bibinfo {author} {\bibfnamefont {P.~L.}\ \bibnamefont {Morton}}, \
    and\ \bibinfo {author} {\bibfnamefont {M.~N.}\ \bibnamefont {Rosenbluth}},\
    }\href {\doibase 10.1109/JQE.1981.1071285} {\bibfield  {journal} {\bibinfo
    {journal} {IEEE J. Quantum Electron.}\ }\textbf {\bibinfo {volume} {17}},\
    \bibinfo {pages} {1436} (\bibinfo {year} {1981})}\BibitemShut {NoStop}%
  \bibitem [{\citenamefont {Emma}\ \emph {et~al.}(2017)\citenamefont {Emma},
    \citenamefont {Sudar}, \citenamefont {Musumeci}, \citenamefont {Urbanowicz},\
    and\ \citenamefont {Pellegrini}}]{Emma.2017}%
    \BibitemOpen
    \bibfield  {author} {\bibinfo {author} {\bibfnamefont {C.}~\bibnamefont
    {Emma}}, \bibinfo {author} {\bibfnamefont {N.}~\bibnamefont {Sudar}},
    \bibinfo {author} {\bibfnamefont {P.}~\bibnamefont {Musumeci}}, \bibinfo
    {author} {\bibfnamefont {A.}~\bibnamefont {Urbanowicz}}, \ and\ \bibinfo
    {author} {\bibfnamefont {C.}~\bibnamefont {Pellegrini}},\ }\href {\doibase
    10.1103/PhysRevAccelBeams.20.110701} {\bibfield  {journal} {\bibinfo
    {journal} {Phys. Rev. Accel. Beams}\ }\textbf {\bibinfo {volume} {20}}
    (\bibinfo {year} {2017}),\ 10.1103/PhysRevAccelBeams.20.110701}\BibitemShut
    {NoStop}%
  \bibitem [{\citenamefont {Prosnitz}\ \emph {et~al.}(1981)\citenamefont
    {Prosnitz}, \citenamefont {Szoke},\ and\ \citenamefont
    {Neil}}]{Prosnitz.1981}%
    \BibitemOpen
    \bibfield  {author} {\bibinfo {author} {\bibfnamefont {D.}~\bibnamefont
    {Prosnitz}}, \bibinfo {author} {\bibfnamefont {A.}~\bibnamefont {Szoke}}, \
    and\ \bibinfo {author} {\bibfnamefont {V.~K.}\ \bibnamefont {Neil}},\ }\href
    {\doibase 10.1103/PhysRevA.24.1436} {\bibfield  {journal} {\bibinfo
    {journal} {Phys. Rev. A}\ }\textbf {\bibinfo {volume} {24}},\ \bibinfo
    {pages} {1436} (\bibinfo {year} {1981})}\BibitemShut {NoStop}%
  \bibitem [{\citenamefont {Mak}\ \emph {et~al.}(2017)\citenamefont {Mak},
    \citenamefont {Curbis},\ and\ \citenamefont {Werin}}]{Mak.2017}%
    \BibitemOpen
    \bibfield  {author} {\bibinfo {author} {\bibfnamefont {A.}~\bibnamefont
    {Mak}}, \bibinfo {author} {\bibfnamefont {F.}~\bibnamefont {Curbis}}, \ and\
    \bibinfo {author} {\bibfnamefont {S.}~\bibnamefont {Werin}},\ }\href
    {\doibase 10.1103/PhysRevAccelBeams.20.060703} {\bibfield  {journal}
    {\bibinfo  {journal} {Phys. Rev. Accel. Beams}\ }\textbf {\bibinfo {volume}
    {20}},\ \bibinfo {pages} {60703} (\bibinfo {year} {2017})},\ \Eprint
    {http://arxiv.org/abs/1611.04925} {arXiv:1611.04925} \BibitemShut {NoStop}%
  \bibitem [{\citenamefont {Mak}\ \emph {et~al.}(2015)\citenamefont {Mak},
    \citenamefont {Curbis},\ and\ \citenamefont {Werin}}]{Mak.2015}%
    \BibitemOpen
    \bibfield  {author} {\bibinfo {author} {\bibfnamefont {A.}~\bibnamefont
    {Mak}}, \bibinfo {author} {\bibfnamefont {F.}~\bibnamefont {Curbis}}, \ and\
    \bibinfo {author} {\bibfnamefont {S.}~\bibnamefont {Werin}},\ }\href
    {\doibase 10.1103/PhysRevSTAB.18.040702} {\bibfield  {journal} {\bibinfo
    {journal} {Phys. Rev. Spec. Top. - Accel. Beams}\ }\textbf {\bibinfo {volume}
    {18}} (\bibinfo {year} {2015}),\ 10.1103/PhysRevSTAB.18.040702}\BibitemShut
    {NoStop}%
  \bibitem [{\citenamefont {Born}\ \emph {et~al.}(1999)\citenamefont {Born},
    \citenamefont {Wolf}, \citenamefont {Bhatia}, \citenamefont {Clemmow},
    \citenamefont {Gabor}, \citenamefont {Stokes}, \citenamefont {Taylor},
    \citenamefont {Wayman},\ and\ \citenamefont
    {Wilcock}}]{born_wolf_bhatia_clemmow_gabor_stokes_taylor_wayman_wilcock_1999}%
    \BibitemOpen
    \bibfield  {author} {\bibinfo {author} {\bibfnamefont {M.}~\bibnamefont
    {Born}}, \bibinfo {author} {\bibfnamefont {E.}~\bibnamefont {Wolf}}, \bibinfo
    {author} {\bibfnamefont {A.~B.}\ \bibnamefont {Bhatia}}, \bibinfo {author}
    {\bibfnamefont {P.~C.}\ \bibnamefont {Clemmow}}, \bibinfo {author}
    {\bibfnamefont {D.}~\bibnamefont {Gabor}}, \bibinfo {author} {\bibfnamefont
    {A.~R.}\ \bibnamefont {Stokes}}, \bibinfo {author} {\bibfnamefont {A.~M.}\
    \bibnamefont {Taylor}}, \bibinfo {author} {\bibfnamefont {P.~A.}\
    \bibnamefont {Wayman}}, \ and\ \bibinfo {author} {\bibfnamefont {W.~L.}\
    \bibnamefont {Wilcock}},\ }\href {\doibase 10.1017/cbo9781139644181} {\emph
    {\bibinfo {title} {Princ. Opt.}}},\ \bibinfo {edition} {7th}\ ed.\ (\bibinfo
    {publisher} {Cambridge University Press},\ \bibinfo {year}
    {1999})\BibitemShut {NoStop}%
  \bibitem [{\citenamefont {Reiche}(1999)}]{reiche1999genesis}%
    \BibitemOpen
    \bibfield  {author} {\bibinfo {author} {\bibfnamefont {S.}~\bibnamefont
    {Reiche}},\ }\href {\doibase 10.1016/S0168-9002(99)00114-X} {\bibfield
    {journal} {\bibinfo  {journal} {Nucl. Instruments Methods Phys. Res. Sect. A
    Accel. Spectrometers, Detect. Assoc. Equip.}\ }\textbf {\bibinfo {volume}
    {429}},\ \bibinfo {pages} {243} (\bibinfo {year} {1999})}\BibitemShut
    {NoStop}%
  \bibitem [{\citenamefont {Karssenberg}\ \emph {et~al.}(2006)\citenamefont
    {Karssenberg}, \citenamefont {{Van Der Slot}}, \citenamefont {Volokhine},
    \citenamefont {Verschuur},\ and\ \citenamefont {Boller}}]{Karssenberg.2006}%
    \BibitemOpen
    \bibfield  {author} {\bibinfo {author} {\bibfnamefont {J.~G.}\ \bibnamefont
    {Karssenberg}}, \bibinfo {author} {\bibfnamefont {P.~J.}\ \bibnamefont {{Van
    Der Slot}}}, \bibinfo {author} {\bibfnamefont {I.~V.}\ \bibnamefont
    {Volokhine}}, \bibinfo {author} {\bibfnamefont {J.~W.}\ \bibnamefont
    {Verschuur}}, \ and\ \bibinfo {author} {\bibfnamefont {K.~J.}\ \bibnamefont
    {Boller}},\ }\href {\doibase 10.1063/1.2363253} {\bibfield  {journal}
    {\bibinfo  {journal} {J. Appl. Phys.}\ }\textbf {\bibinfo {volume} {100}},\
    \bibinfo {pages} {93106} (\bibinfo {year} {2006})}\BibitemShut {NoStop}%
  \bibitem [{\citenamefont {Huang}\ \emph {et~al.}(2019)\citenamefont {Huang},
    \citenamefont {Li},\ and\ \citenamefont {Deng}}]{Huang.2019}%
    \BibitemOpen
    \bibfield  {author} {\bibinfo {author} {\bibfnamefont {N.~S.}\ \bibnamefont
    {Huang}}, \bibinfo {author} {\bibfnamefont {K.}~\bibnamefont {Li}}, \ and\
    \bibinfo {author} {\bibfnamefont {H.~X.}\ \bibnamefont {Deng}},\ }\href
    {\doibase 10.1007/s41365-019-0559-5} {\bibfield  {journal} {\bibinfo
    {journal} {Nucl. Sci. Tech.}\ }\textbf {\bibinfo {volume} {30}},\ \bibinfo
    {pages} {39} (\bibinfo {year} {2019})}\BibitemShut {NoStop}%
  \bibitem [{\citenamefont {Lindberg}\ \emph {et~al.}(2011)\citenamefont
    {Lindberg}, \citenamefont {Kim}, \citenamefont {Shvyd'Ko},\ and\
    \citenamefont {Fawley}}]{Lindberg.2011}%
    \BibitemOpen
    \bibfield  {author} {\bibinfo {author} {\bibfnamefont {R.~R.}\ \bibnamefont
    {Lindberg}}, \bibinfo {author} {\bibfnamefont {K.~J.}\ \bibnamefont {Kim}},
    \bibinfo {author} {\bibfnamefont {Y.}~\bibnamefont {Shvyd'Ko}}, \ and\
    \bibinfo {author} {\bibfnamefont {W.~M.}\ \bibnamefont {Fawley}},\ }\href
    {\doibase 10.1103/PhysRevSTAB.14.010701} {\bibfield  {journal} {\bibinfo
    {journal} {Phys. Rev. Spec. Top. - Accel. Beams}\ }\textbf {\bibinfo {volume}
    {14}},\ \bibinfo {pages} {10701} (\bibinfo {year} {2011})}\BibitemShut
    {NoStop}%
  \bibitem [{\citenamefont {Ishikawa}(1989)}]{Ishikawa.1989}%
    \BibitemOpen
    \bibfield  {author} {\bibinfo {author} {\bibfnamefont {T.}~\bibnamefont
    {Ishikawa}},\ }\href {\doibase 10.1063/1.1140825} {\bibfield  {journal}
    {\bibinfo  {journal} {Rev. Sci. Instrum.}\ }\textbf {\bibinfo {volume}
    {60}},\ \bibinfo {pages} {2058} (\bibinfo {year} {1989})}\BibitemShut
    {NoStop}%
  \bibitem [{\citenamefont {Malgrange}\ \emph {et~al.}(1991)\citenamefont
    {Malgrange}, \citenamefont {Carvalho}, \citenamefont {Braicovich},\ and\
    \citenamefont {Goulon}}]{Malgrange.1991}%
    \BibitemOpen
    \bibfield  {author} {\bibinfo {author} {\bibfnamefont {C.}~\bibnamefont
    {Malgrange}}, \bibinfo {author} {\bibfnamefont {C.}~\bibnamefont {Carvalho}},
    \bibinfo {author} {\bibfnamefont {L.}~\bibnamefont {Braicovich}}, \ and\
    \bibinfo {author} {\bibfnamefont {J.}~\bibnamefont {Goulon}},\ }\href
    {\doibase 10.1016/0168-9002(91)90676-H} {\bibfield  {journal} {\bibinfo
    {journal} {Nucl. Inst. Methods Phys. Res. A}\ }\textbf {\bibinfo {volume}
    {308}},\ \bibinfo {pages} {390} (\bibinfo {year} {1991})}\BibitemShut
    {NoStop}%
  \bibitem [{\citenamefont {Shvyd'ko}(2019)}]{Shvydko2019}%
    \BibitemOpen
    \bibfield  {author} {\bibinfo {author} {\bibfnamefont {Y.}~\bibnamefont
    {Shvyd'ko}},\ }\href {\doibase 10.1103/PhysRevAccelBeams.22.100703}
    {\bibfield  {journal} {\bibinfo  {journal} {Phys. Rev. Accel. Beams}\
    }\textbf {\bibinfo {volume} {22}} (\bibinfo {year} {2019}),\
    10.1103/PhysRevAccelBeams.22.100703}\BibitemShut {NoStop}%
  \bibitem [{\citenamefont {Shvyd'Ko}\ and\ \citenamefont
    {Lindberg}(2012)}]{ShvydKo2012}%
    \BibitemOpen
    \bibfield  {author} {\bibinfo {author} {\bibfnamefont {Y.}~\bibnamefont
    {Shvyd'Ko}}\ and\ \bibinfo {author} {\bibfnamefont {R.}~\bibnamefont
    {Lindberg}},\ }\href {\doibase 10.1103/PhysRevSTAB.15.100702} {\bibfield
    {journal} {\bibinfo  {journal} {Phys. Rev. Spec. Top. - Accel. Beams}\
    }\textbf {\bibinfo {volume} {15}},\ \bibinfo {pages} {1} (\bibinfo {year}
    {2012})},\ \Eprint {http://arxiv.org/abs/arXiv:1207.3376v1}
    {arXiv:arXiv:1207.3376v1} \BibitemShut {NoStop}%
  \end{thebibliography}

%

\end{document}